\newif\ifconfver
\confverfalse      

\newif\ifcutshort      
\cutshorttrue

\newif\ifcutshortlvltwo  
\cutshortlvltwofalse

\ifconfver
    \documentclass[10pt,twocolumn,twoside]{IEEEtran}
\else
    \documentclass{IEEEtran}
\fi

\usepackage{cite}
\usepackage{graphicx}
\usepackage{psfrag}
\usepackage{url}
\usepackage{stfloats}
\usepackage{amsfonts,amssymb,amsmath,bm,paralist,theorem,cite,ifthen}
\usepackage{array}
\usepackage{caption}
\usepackage{cases}
\usepackage{calc}

\usepackage{longtable}
\usepackage{stfloats}  
\usepackage{graphics,booktabs,color,epsfig,enumerate}
\usepackage{multirow}
\usepackage{algorithm,algpseudocode}
\usepackage{subfigure}
\usepackage{epstopdf}
\usepackage{array}
\usepackage{stfloats}

\theorembodyfont{\rmfamily}

{\begin{list}{}{
    \settowidth{\labelwidth}{\mbox{\textnormal{#1}}}%
    \setlength{\leftmargin}{\labelwidth+\labelsep}}}%
{\end{list}}

\makeatletter
\def\changeBibColor#1{%
  \in@{#1}{}
  \ifin@\color{blue}\else\normalcolor\fi
}

\begin{document}

\bibliographystyle{IEEEtran}

\newcommand\bcc[2][c]{\ensuremath{\bm{\mathcal{#2}}}}      
\newcommand\bcl[2][c]{\ensuremath{\bm{#2}}}
\newcommand\Ib{\ensuremath{{\bm I}}}
\newcommand\Vb{\ensuremath{{\bm V}}}
\newcommand\vb{\ensuremath{{\bm v}}}
\newcommand\Hb{\ensuremath{{\bm H}}}
\newcommand\ub{\ensuremath{{\bm u}}}
\newcommand\hb{\ensuremath{{\bm h}}}
\newcommand\xb{\ensuremath{{\bm x}}}
\newcommand\gb{\ensuremath{{\bm g}}}
\newcommand\Thetab{\ensuremath{{\bm \Theta}}}
\newcommand\thetab{\ensuremath{{\boldsymbol \theta}}}
\newcommand\Gb{\ensuremath{{\bm G}}}
\newcommand\Pb{\ensuremath{{\bm P}}}
\newcommand\taub{\ensuremath{{\boldsymbol \tau}}}
\newcommand\zb{\ensuremath{{\bm z}}}
\newcommand\st{\ensuremath{{\rm ~s.t.}}}


\def\blue{\color{blue}}
\def\red{\color{red}}
\definecolor{orange}{RGB}{255,107,0}
\def\orange{\color{orange}}

\title{One-Bit Channel Estimation for IRS-aided Millimeter-Wave  Massive MU-MISO System}

\ifconfver \else {\linespread{1.1} \rm \fi

\author{ Silei Wang,  Qiang Li and Jingran Lin
	\thanks{S. Wang, Q. Li ({\it Corresponding author}) and J. Lin are with
		School of Information and Communication  Engineering, University of Electronic Science and Technology of China, Chengdu, P.~R.~China, 611731.
		 E-mails: wangniuqing520@gmail.com, lq@uestc.edu.cn, jingranlin@uestc.edu.cn.
	 
 This work was supported by the National Natural Science
 Foundation of China under Grants 62171110. }
}
\maketitle

\begin{abstract}

Recently, intelligent reflecting surface (IRS)-assisted communication  has gained considerable attention due to its advantage in extending the coverage and compensating the path loss with low-cost passive metasurface.
This  paper considers the uplink channel estimation for IRS-aided multiuser massive multi-input single-output (MISO) communications with  one-bit ADCs  at the base station (BS). The use of one-bit  ADC is  impelled  by the low-cost and power efficient implementation of massive antennas techniques. However, the passiveness of IRS and the lack of signal level information after one-bit quantization make the IRS channel estimation challenging. To tackle this problem, we exploit the  structured sparsity of the user-IRS-BS cascaded channels and develop three channel estimators, each of which utilizes the  structured sparsity at different levels. Specifically, the first estimator exploits the elementwise sparsity of the cascaded channel and employs the sparse Bayesian learning (SBL)  to infer the channel responses via the type-II maximum likelihood (ML) estimation. However, due to the one-bit quantization, the type-II ML in general is intractable. As such, a variational expectation-maximization (EM) algorithm is custom-derived to iteratively compute an ML solution. The second estimator utilizes the common row-structured sparsity induced by the IRS-to-BS channel shared among the users, and develops another type-II ML solution via the block SBL (BSBL) and the variational EM. To further improve the performance of BSBL, a third two-stage estimator is proposed, which can utilize both the common row-structured sparsity and the column-structured sparsity arising from the limited scattering around the users. Simulation results show that the more diverse structured sparsity is exploited, the better estimation performance is achieved, and that the proposed estimators are superior to  state-of-the-art one-bit estimators.


\end{abstract}


\ifconfver \else \IEEEpeerreviewmaketitle} \fi

\section{Introduction}
Low-cost, energy efficient and high spectral efficiency  transmission techniques are indispensable for 6G communications. Recently, the intelligent reflecting surface (IRS) has been put forward and gained considerable attention. The idea of IRS is to deploy a large planar array  to intentionally create  additional  reflecting paths for the receivers. More specifically, the IRS consists of a large number of passive  elements, each of which is capable of independently reflecting the incident electromagnetic wave with certain  phase shifts. By intelligently adjusting the phase shifts, IRS is able  to adapt to wireless channels and create a favorable scattering environment around the transceiver. As compared with the relay technique, IRS can be easily implemented with passive and low-cost diodes with low power consumption, which is attractive for the next generation communications~\cite{survey20,di2020smart}.

To fully exploits the benefits of the IRS, it is crucial to jointly design the transmit signal and the IRS phase shift, so that the reflected signals can be constructively (destructively) aligned at the desired (non-intended) receivers~\cite{wu2019towards,liaskos2018new}. Therefore, a vast majority of works on IRS have focused on the joint design for IRS-aided communication systems, such as multi-input multi-output (MIMO) communications~\cite{wu2019intelligent,Cui23}, simultaneous wireless information and power transfer~\cite{zargari2021energy,swipt-Wu22,SWIP-IRS-1}, physical-layer security~\cite{feng2021physical,Secure-1,Secure-0,Secure-2}  and integrated sensing and communication (ISAC)~\cite{ISAC-1,ISAC-2,ISAC-3}, to name a few. It should be pointed out that all the above works have made a blanket assumption that the  channel state information (CSI) of the IRS  links are known as a prior. However, due to the passiveness of the IRS, it could be challenging to acquire the CSI by using the traditional active training methods. In view of this,  there have been substantial works to address the passive channel estimation problem; see the recent survey paper~\cite{survey22}. An on-off strategy was proposed in~\cite{mishra2019channel}, where the IRS turns on each reflecting element sequentially, while keeping the rest reflecting elements closed. Then, the base station (BS) sequentially estimates the channels from the BS to one reflecting element and from this element to the users. The works~\cite{tensor_irs21,tensor-22,tensor-CL}  employed the parallel factor (PARAFAC) tensor model to estimate the transmitter-to-IRS and the IRS-to-receiver MIMO channels, respectively (resp.). Unlike~\cite{tensor_irs21,tensor-22,tensor-CL},  it was advocated in~\cite{mishra2019channel} to directly estimate the transmitter-IRS-receiver cascaded channel, because the subsequent transmit optimization depends  on the cascaded channel. In light of this, the authors in~\cite{zheng2019intelligent,Zhangrui21} designed some novel phase-shift patterns to facilitate the cascaded channel estimation. In~\cite{wang2020compressed, wei2021channel,sparse1,sparse2,sparse3}, the authors proposed to estimate the cascaded channel from a sparse signal recovery perspective by exploiting the channel sparsity in the angular domain for millimeter wave (mmWave) communications.  More recently, the deep learning-based channel estimation approach has also gained much attentions~\cite{DL21,DL23-1,DL23-2}.

In this work, we consider the uplink channel estimation  for IRS-aided mmWave massive multiuser multi-input single-output  (MU-MISO) communications, with an emphasis on the \emph{one-bit} quantization of the received signals at the BS, i.e., each antenna of the BS is equipped with a pair of one-bit analog-to-digital converters (ADCs) for I/Q channels. The use of one-bit  ADC is  impelled  by the low-cost and energy efficient implementation of massive MIMO. It is known that the power consumption of ADCs scales roughly exponentially with the number of quantization bits~\cite{Svensson2006On}, and thus using one-bit ADC can largely save the power consumption especially when the number of antennas is large, as in the massive MIMO case. In addition, the one-bit ADC has constant envelop or ideal peak-to-average-power ratio (PAPR), which allows to use low-cost and high power-efficiency amplifier to transmit and receive. Despite these attractive features of one-bit massive MIMO, the lack of signal level information also poses difficulty in channel estimation. There have been some endeavors trying to address this problem in the conventional MIMO communications without IRS~\cite{Choi2016near,li2017channel,nguyen2021svm,mo2014channel,stockle2016channel, zhou2022millimeter}. In~\cite{Choi2016near}, a near maximum likelihood (nML) channel estimator and detector was proposed. The Bussgang decomposition was exploited in~\cite{li2017channel} to develop a Bussgang-based linear minimum mean-squared error (BLMMSE) channel estimator for massive MIMO systems with one-bit ADCs. A channel estimation scheme based on support vector machine (SVM) with one-bit ADCs was presented in~\cite{nguyen2021svm}. By taking into account the sparsity of the millimeter-wave MIMO channels, the estimation problem is formulated as a one-bit compressed sensing problem, and the expectation-maximization (EM) algorithm has been applied for maximum likelihood (ML) estimation and maximum a posteriori (MAP) estimation from one-bit measurements in~\cite{mo2014channel} and~\cite{stockle2016channel}, resp. The authors in~\cite{zhou2022millimeter} exploited the sparse and low-rank properties of the mmWave channel with one-bit ADCs to develop a mmWave channel estimation algorithm.
Although various one-bit channel estimation algorithms have been proposed for the conventional MIMO communication systems, directly applying them to the IRS channels usually leads to  performance degradation, because of the mismatch between the algorithms' prerequisites and the characteristics of the IRS channels; we will elaborate on this in the numerical results Sec.~\ref{sec:num_results}. In addition, the shared IRS-to-BS channel among the users makes IRS channels some distinct structure, which in general does not exhibit in the conventional MIMO channels. Therefore, by carefully exploiting the structure of the IRS  channels, it is possible to custom-derive   better estimation algorithms rather than just  adapting the existing  one-bit MIMO channel  estimation algorithms~\cite{Choi2016near,li2017channel,nguyen2021svm,mo2014channel,stockle2016channel, zhou2022millimeter} to the IRS channels. To this end, several estimation algorithms are proposed in this work with different levels of utilization of the IRS channel structures.

By exploiting the sparsity of the mmWave channels in the angular domain, we first establish a sparse signal recovery model for the IRS channel estimation, and employ  the hierarchical sparse Bayesian learning (SBL)~\cite{Tipping2001Sparse} to adaptively infer the channels. Specifically, by imposing a parameterized  sparsity-promoting prior on the (angular domain) cascaded channels, the SBL aims at estimating the hyperparameters governing the prior via type-II ML, i.e.,
integrating out the unquantized receive signals and the angular channel responses, and then performing ML optimization.  However, the type-II ML  problem is challenging since the  marginal likelihood under the one-bit quantization  has no analytic form. To circumvent this difficulty,  we propose a variational EM algorithm, which combines the EM algorithm with the mean-field variational inference~\cite{hoffman2013stochastic}, to simultaneously estimate the  hyperparameters as well as  infer the channels. Compared to the EM-BPDN algorithm in~\cite{stockle2016channel}, which treats the channel as unknown parameter with a fixed sparsity-promoting prior, the proposed  hierarchical SBL method can not only achieve automatic model selection, but also prevent any structural errors~\cite{wipf2004sparse}.

We should point out that the above SBL estimator exploits the sparsity of the mmWave channels for each user without considering the correlation among them. However, as mentioned before, since all the users share the  IRS-to-BS channel, this brings about additional structure among the users' channels.  As we will see in Sec.~\ref{sec:BSBL}, considering limited scattering around the BS, the shared IRS-to-BS channel translates to common row-structured sparsity shared by all the users' (angular domain) cascaded channels. In light of this, we formulate the channel estimation problem a block sparse signal recovery problem, and derive a block SBL (BSBL)~\cite{zhang2011sparse} estimator by again using the variational EM approach. Numerical results show that the BSBL estimator has better performance than the SBL estimator.


While the BSBL approach takes advantages of the common row-structured sparsity of the angular cascaded channels, it is unable to capture the sparsity within the non-zero blocks due to the modeling limitation. To be specific, as we will see in Sec.~\ref{sec:two_stage}, with  limited scattering  around the IRS, the non-zero rows of the angular cascaded channels should also possess some sparsity. To fully exploit this double-level sparsity, we  further propose a heuristic two-stage SBL-based  estimation scheme to improve
the performance of the BSBL scheme. Specifically, in the first stage, we perform the BSBL scheme to estimate the common row support of the angular cascaded channels. In the second stage, a refined channel estimation based on the SBL scheme is performed within the non-zero row support found in the first stage. Since this two-stage scheme exploits both the row-structured sparsity and the column-structured sparsity of the angular cascaded channels, it can effectively reduce the estimation error as compared with the previous two.

We consider an IRS-aided multiuser MISO channel estimation with one-bit quantized measurements at the BS. Our main contributions are summarized as follows.
\begin{enumerate}
\item By exploiting the limited scattering property of mmWave communications and the cascaded channel structure induced by IRS, we formulate the channel estimation as a structured sparse signal recovery problem. Different levels of utilization of  the  structured sparsity are investigated, including elementwise-level sparsity, row-level sparsity and row-column-level sparsity.

\item  By considering different levels of  structured sparsity, three SBL-based channel estimation schemes are proposed, namely the SBL scheme, the BSBL scheme and the two-stage SBL scheme.  Different from the classical SBL~\cite{wipf2004sparse} and Block SBL algorithms~\cite{zhang2011sparse}, which are developed based  on a linear observation model, our proposed algorithms handle nonlinear one-bit quantization model, under which the conventional EM method in~\cite{wipf2004sparse,zhang2011sparse} is no longer applicable due to  the intractable  
joint posterior distribution of hidden variables. To circumvent this difficulty, we propose a new approach based on the idea of variational mean-field inference~\cite{hoffman2013stochastic} and block-coordinate ascent (BCA) optimization to iteratively approximate the joint posterior distribution. We show that despite the presence of one-bit quantization, the resulting Bayesian estimation problems can be still handled under the EM framework with efficient closed-form update. 


\item To enhance the proposed algorithms' efficiency, we introduce a virtual angular domain (VAD)-dictionary-based scheme for designing IRS phase-shifts. This scheme takes advantage of the unique hyperparameter updating structure in the proposed SBL algorithms. Theoretical analysis and numerical simulations confirm the substantial complexity reduction achieved by our approach.


\end{enumerate}

\subsection{Organization and Notations}
The remainder of this paper is organized as follows. Section~\ref{sec:system_model} introduces the system model and  the sparse structure of the cascaded channels. Section~\ref{sec:SBL} studies the SBL estimation scheme. Section~\ref{sec:BSBL} studies the BSBL  scheme. A two-stage estimation scheme is developed in Section~\ref{sec:two_stage}. The complexity analysis and the fast implementation of the SBL estimator are provided in Section~\ref{sec:complexity_fastSBL}. The simulation results are provided in Section~\ref{sec:num_results}, and Section~\ref{sec:conclusions} concludes the paper.

Our notations are as follows. Upper (lower) bold face letters are used for matrices (vectors); $(\cdot)^T$ and $(\cdot)^H$ denote transpose and Hermitian transpose, resp. $\text{Tr}(\cdot)$ denotes a trace operation. $\otimes$ stands for the Kronecker product. $\mathbf{I}$, $\bm 0$ and $\bm 1$ denote the identity matrix, the all-zero matrix/vector and  the all-one vector with appropriate dimension, resp. $\| \cdot \|$ and $\| \cdot\|_1$ denote the Euclidean norm and $\ell_1$-norm, resp. $\text{Diag}(\bm x)$ denotes a diagonal matrix with  diagonal elements being  $\bm x$. $\mathfrak{R}\{z\}$ and $\mathfrak{I}\{z\}$ denote the real and imaginary part of a complex number $z$, resp. $\mathbb{C}$, $\mathbb{R}$ and $\mathbb{R}_+$ denote the one-dimensional space of complex, real and nonnegative real numbers, resp. $\mathbb{H}_{++}^n$ denotes the set of $n$-by-$n$ positive definite matrices; $|\mathcal{S}|$ denotes the number of elements in set $\mathcal{S}$. $\bm A^{\dag}$ is the Moore-Penrose pseudoinverse matrix. $\text{det}(\bm A)$ and $\|\bm A\|_F$ denote the determinant and the Frobenius norm of matrix $\bm A$, resp. $\text{Vec}(\bm A)$ denotes a vector formed by stacking $\bm A$ column by column and $\text{Vec}^{-1}(\bm a)$ denotes an  inverse  operation of $\text{Vec}(\cdot)$, i.e. $ \text{Vec}^{-1}(\bm a)= \bm A$ if $\bm a =  \text{Vec}(\bm A)$. $\text{Diag}({\bm A}, {\bm B})$ represents a block diagonal matrix with the diagonal blocks being $\bm A$ and $\bm B$. $[\bm x]_{a:b}$ denotes a subvector formed from the $a$-th element to the $b$-th element of $\bm x$. $[\bm A]_{a:b,c:d}$ denotes a sub-matrix of $\bm A$ with rows and columns  indexed from $a$ to $b$ and $c$ to $d$, resp. Similarly, $[\bm A]_{:,c:d}$ denotes a sub-matrix of $\bm A$ with columns indexed from  $c$ to $d$. $[\bm A]_{:,\bm \Omega}$ and $[\bm A]_{\bm \Omega,:}$ denote the reduced matrices consisting of the columns and the rows of $\bm A$ indexed by the elements of the set $\bm \Omega$, resp. $\mathcal{CN} ({\bm x}, {\bm \Sigma})$ denotes the complex Gaussian distribution with mean $\bm x$ and covariance  $\bm \Sigma$. Statistical expectation is represented by $\mathbb{E}[\cdot]$.


\section{System Model and Prior Work} \label{sec:system_model}
\begin{figure}[!h]
	\centerline{\resizebox{.4\textwidth}{!}{\includegraphics{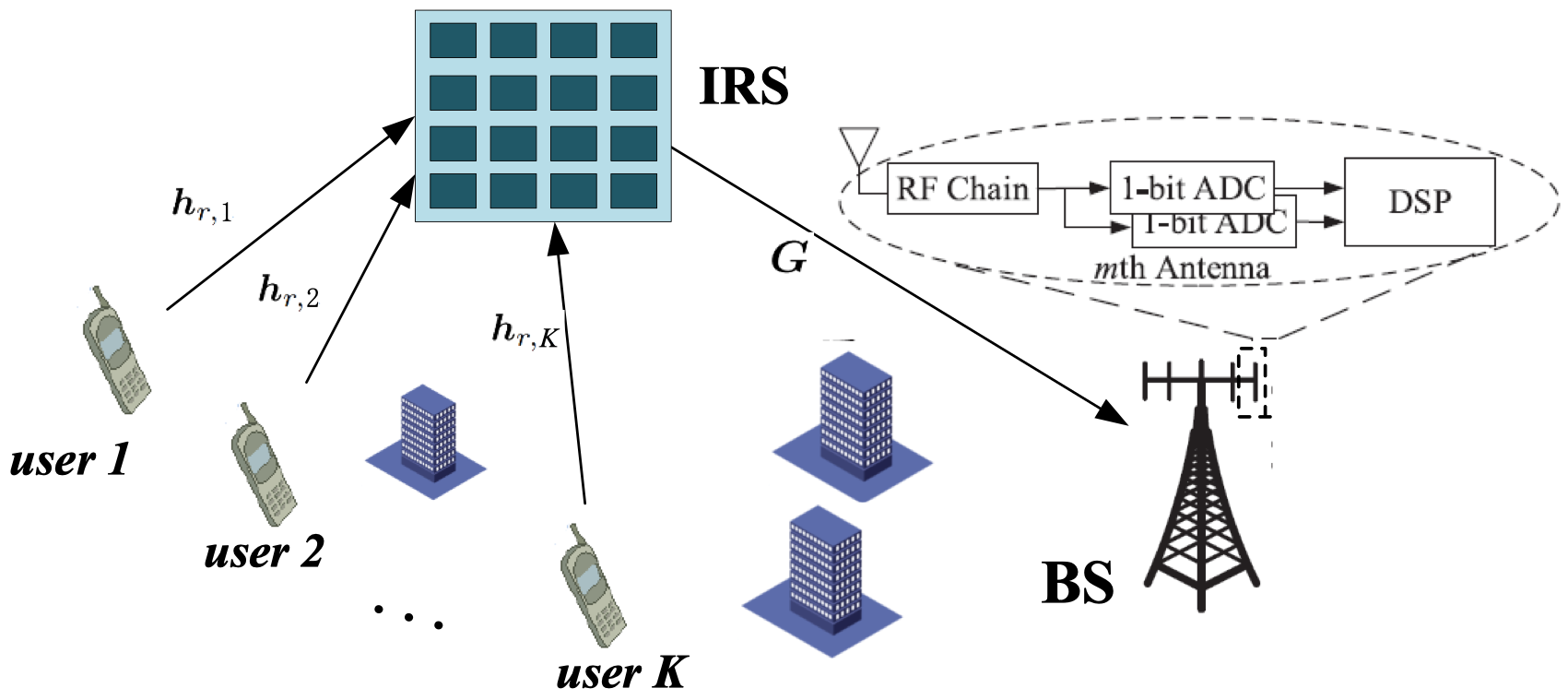}}
	}   \caption{An IRS-aided  MU-MISO mmWave communication system.} \label{fig:system_model}
\end{figure}

\begin{figure}[!h]
	\centerline{\resizebox{.4\textwidth}{!}{\includegraphics{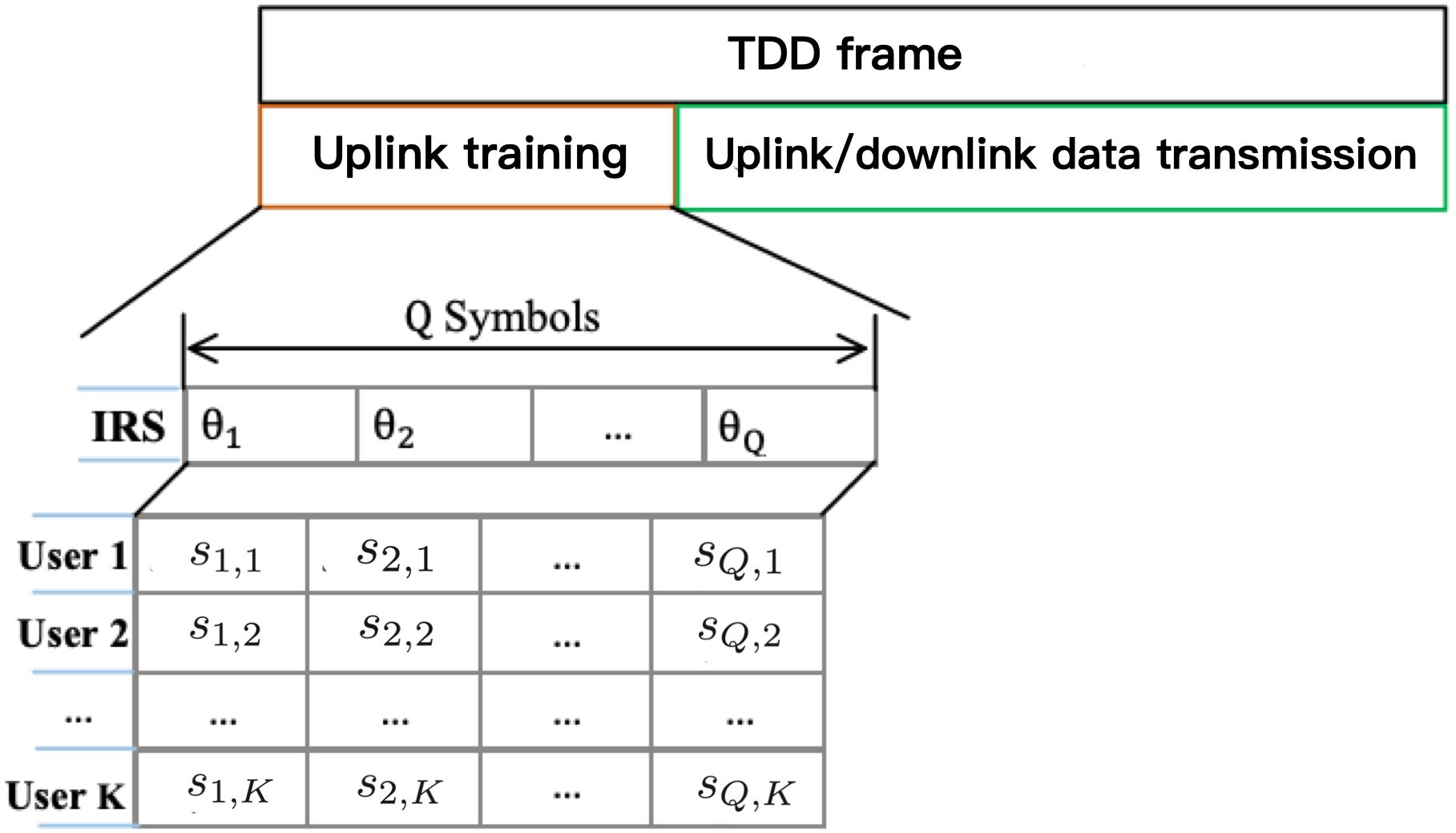}}
	}   \caption{The TDD-based frame structure.} \label{fig:Frame_structure}
\end{figure}

As shown in Fig.~\ref{fig:system_model}, we consider an MU-MISO mmWave communication system operating in time division duplex (TDD) mode, where  $K$ single-antenna users communicate with the BS through  the IRS, and for simplicity, direct link is not considered between the users and the BS\footnote{It is possible to include the direct link in the subsequent development; we  discuss this in the supplementary material of this paper.}.
The BS is a uniform linear array (ULA) with $M$ receive antennas, each of which is equipped with a pair of one-bit ADCs for I/Q channels. The IRS is a uniform planar array (UPA) with $N$ reflecting elements.
We consider a TDD based frame structure as shown in Fig.~\ref{fig:Frame_structure}. The frame structure is divided into two phases: the first phase is for the uplink training and the second phase is for uplink/downlink data transmission. In this work, we only focus on the uplink training. Specifically, during the uplink training, $K$ users simultaneously transmit pilot symbols to the BS via the IRS over $Q$ time slots. Denote the transmitted pilot symbol from the $k$-th user in the $q$-th time slot as $s_{q,k}$, $\forall q=1,\ldots,Q$, $\forall k\in \mathcal{K}$. The baseband received signal at the BS in the $q$-th time slot can be represented as
\begin{equation} \label{eq:y_q}
	\bm y_q = \sum_{k=1}^{K} \bm H_k \bm \theta_q s_{q,k} +\bm w_q, ~ q=1,\ldots, Q,
\end{equation}
where $\bm H_k \triangleq \bm G \text{Diag}(\bm h_{r,k})$ is referred to as the cascaded channel for the $k$-th user;  $\bm G\in \mathbb{C}^{M\times N}$ and $\bm h_{r,k}\in \mathbb{C}^{N\times 1}$ $\forall k\in \mathcal{K}\triangleq \{1,\ldots, K\}$ denote the channels from the IRS to the BS and from the $k$-th user to the IRS, resp.; $\bm \theta_q=[\theta_{q,1},\ldots,\theta_{q,N}]^T\in \mathbb{C}^{N\times 1}$ denotes the reflecting vector at the IRS with $\theta_{q,n}$ being the reflecting coefficient at the $n$-th IRS reflecting element in the $q$-th time slot; $\bm w_q \in \mathbb{C}^{M\times 1}$ is  noise following $\bm w_q \sim \mathcal{CN}(\bm 0, \sigma^2 \bm I)$. Since in the downlink transmission, the joint active beamforming and IRS phase-shift design depends on $\bm h_{r,k}$ and $\bm G$ through the cascaded channel $\bm H_k$, it suffices to  know   $\bm H_k$, $\forall~k$  for the downlink transmit design~\cite{survey22}. Therefore, in this  paper we will focus on estimating the cascaded channel $\bm H_k$, $\forall~k$.

With the one-bit ADCs, the received signal at the BS after quantization is given by
\begin{equation}
	\bm r_q=\text{sgn}(\bm y_q), ~ q=1,\ldots, Q,
\end{equation}
where $\text{sgn}(\cdot)$ denotes the elementwise sign function and is applied separately to the real and imaginary parts, i.e., $\text{sgn}(y)=\text{sgn}(\mathfrak{R}\{y\})+j \text{sgn}(\mathfrak{I}\{y\})$ with $j=\sqrt{-1}$ and
\begin{numcases}{\text{sgn}(x)=}
-1, ~~~x\leq 0, \notag \\
+1, ~~~x>0.
\end{numcases}
We have $r_{q,i}\in \mathcal{S}\triangleq \{1+j,-1+j,-1-j,1-j\}$ for $i=1,\ldots,M$.
\subsection{Sparse Cascaded Channel Model}
Following the geometric channel model~\cite{hu2018super}, the channels $\bm G$ and $\bm h_{r,k}$ are resp. given by
\begin{equation} \label{eq:G}
	\bm G \hspace{-3pt}=\hspace{-3pt} \frac{1}{\sqrt{L_G}} \hspace{-3pt}\sum_{\ell_1=1}^{L_G}\hspace{-2pt} \zeta_{\ell_1}^G \bm a_M \left(\frac{2 d}{\lambda} \sin (\vartheta_{\ell_1}^{AoA})\right)\bm b_N^H (u_{\ell_1}^{AoD},\nu_{\ell_1}^{AoD})
\end{equation}
\begin{equation} \label{eq:hr}
	\bm h_{r,k} = \frac{1}{\sqrt{L_{r,k}}} \sum_{\ell_2=1}^{L_{r,k}} \zeta_{\ell_2,k}^r \bm b_N ( u_{\ell_2,k}^{AoA}, \nu_{\ell_2,k}^{AoA})
\end{equation}
with
\[
	u_{\ell_1}^{AoD} \triangleq \frac{2d}{\lambda}\sin (\theta_{\ell_1}^{AoD}) \sin (\gamma_{\ell_1}^{AoD} ),~\nu_{\ell_1}^{AoD} \triangleq \frac{2d}{\lambda}\cos (\theta_{\ell_1}^{AoD}),
\]
\[
	u_{\ell_2,k}^{AoA} \triangleq \frac{2d}{\lambda}\sin (\theta_{\ell_2,k}^{AoA}) \sin (\gamma_{\ell_2,k}^{AoA} ),~\nu_{\ell_2,k}^{AoA} \triangleq \frac{2d}{\lambda}\cos (\theta_{\ell_2,k}^{AoA}),
\]
where $\zeta_{\ell_1}^G$ and $\zeta_{\ell_2,k}^r$ denote the complex gains of the $\ell_1$-th spatial path from the IRS to the BS and the $\ell_2$-th spatial path from the $k$-th user to the IRS, resp., and $\zeta_{\ell_1}^G$ and $\zeta_{\ell_2,k}^r$   are assumed to be independent with zero mean;  $\vartheta_{\ell_1}^{AoA}$ is the angle of arrival (AoA) at the BS; $\gamma_{\ell_1}^{AoD}$($\theta_{\ell_1}^{AoD}$) denotes the azimuth (elevation) angle of departure (AoD) at the IRS for the $\ell_1$-th path; $\gamma_{\ell_2,k}^{AoA}$($\theta_{\ell_2,k}^{AoA}$) denotes the azimuth (elevation) AoA for the $\ell_2$-th path from the $k$-th user to the IRS. $L_G$ is the number of paths between the IRS and the BS, and $L_{r,k}$ is the number of paths between the $k$-th user and the IRS. For mmWave communications, considering limited scattering around the BS and the IRS, it typically holds $L_G \ll \min\{M,N\}$ and $L_{r,k}\ll N$. The $d$ and $\lambda$ denote the antenna spacing and the carrier wavelength, resp., and satisfy $d/\lambda=1/2$ for simplicity. For a typical  $N_x\times N_y$ ($N=N_x\times N_y$) UPA, $\bm b_N (u,\nu)$ can be expressed as
\begin{equation}
	\bm b_N (u,\nu) = \bm a_{N_x}(u)\otimes \bm a_{N_y}(\nu),
\end{equation}
where $\bm a_X(\nu)\in \mathbb{C}^{X\times 1}$ is the array steering vector, i.e.,
\vspace{-5pt}
\begin{equation} \label{eq:steer_vec}
	\bm a_X(\nu) = [1, e^{j \pi \nu},\ldots, e^{j \pi \nu (X-1)} ]^H.
\end{equation}
With~\eqref{eq:G} and~\eqref{eq:hr}, the cascaded channel $\bm H_k$  is expressed as
\begin{equation} \label{eq:cascaded_ch}
\begin{aligned}
	\bm H_k &= \frac{1}{\sqrt{L_G L_{r,k}}} \sum_{\ell_1=1}^{L_G} \sum_{\ell_2=1}^{L_{r,k}} \zeta_{\ell_1}^G \zeta_{\ell_2,k}^r \bm a_M \left(\sin (\vartheta_{\ell_1}^{AoA}) \right) \\
	& \times \bm b_N^H (u_{\ell_1}^{AoD}- u_{\ell_2,k}^{AoA},\nu_{\ell_1}^{AoD} - \nu_{\ell_2,k}^{AoA}).
\end{aligned}
\end{equation}
For simplicity, we call $\bm b_N (u_{\ell_1}^{AoD}- u_{\ell_2,k}^{AoA},\nu_{\ell_1}^{AoD} - \nu_{\ell_2,k}^{AoA})$ as the cascaded AoD steering vector.

To formulate the cascaded channel estimation as a sparse signal recovery problem, we utilize the virtual angular-domain (VAD) representation~\cite{sayeed2002deconstructing} to express  $\bm H_k$ as
\begin{equation} \label{eq:VAD_rep}
	\bm H_k =\bm U_R \widetilde{\bm H}_k (\bm U_{T_x}^H \otimes \bm U_{T_y}^H ) \triangleq \bm U_R \widetilde{\bm H}_k \bm U_T^H
\end{equation}
with
\begin{align*}
\bm U_R & =\frac{1}{\sqrt{M}} \big[\bm a_M(-1), \bm a_M(-1+\frac{2}{G_r} ),\ldots, \bm a_M(1-\frac{2}{G_r} ) \big], \\
\bm U_{T_x} &\hspace{-3pt} =\hspace{-4pt} \frac{1}{\sqrt{N_x}}\big[\bm a_{N_x}(-1), \bm a_{N_x}(-1+\frac{2}{G_{t_x}} ),\ldots, \bm a_{N_x}(1-\frac{2}{G_{t_x}} ) \big]. \\
\bm U_{T_y} &\hspace{-3pt} =\hspace{-4pt} \frac{1}{\sqrt{N_y}} \big[\bm a_{N_y}(-1), \bm a_{N_y}(-1+\frac{2}{G_{t_y}} ),\ldots, \bm a_{N_y}(1-\frac{2}{G_{t_y}} ) \big].
\end{align*}
where $\bm U_R \in \mathbb{C}^{M\times G_r}$ is the VAD transformation dictionary ($G_r\geq M$) with angular resolutions $G_r$ and each column of $\bm U_R$ represents a steering vector parameterized by a pre-discretized angle grid. $\bm U_{T_x} \in \mathbb{C}^{N_x\times G_{t_x}}$ and $\bm U_{T_y} \in \mathbb{C}^{N_y\times G_{t_y}}$ are similarly defined. Therefore, $\bm U_T \triangleq \bm U_{T_x} \otimes \bm U_{T_y}\in \mathbb{C}^{N\times G_t}$ ($G_t = G_{t_x}\times G_{t_y}$) can be regarded as the VAD transformation dictionary ($G_t\geq N$) with angular resolutions $G_t$. Besides, $\widetilde{\bm H}_k\in \mathbb{C}^{G_r \times G_t}$ is the angular domain sparse channel matrix with the non-zero $(i,j)$-th component corresponding to the path gain on the channel consisting of the $i$-th  AoA array steering vector and the $j$-th cascaded AoD array steering vector. Similar to~\cite{wang2020compressed, wei2021channel}, we assume that the true AoA and AoD parameters lie on the discretized grid. With this assumption, $\widetilde{\bm H}_k\in \mathbb{C}^{G_r \times G_t}$ is perfectly sparse, i.e., has $L_G$ non-zero rows, and each non-zero row has $L_{r,k}$ non-zero columns.


\subsection{Problem Formulation}
By substituting the VAD representation $\bm H_k = \bm U_R \widetilde{\bm H}_k \bm U_T^H$ into~\eqref{eq:y_q}, we obtain
\begin{equation} \label{eq:y_q2}
\begin{aligned}
	\bm y_q & = \bm U_R \big(\sum_{k=1}^{K} \widetilde{\bm H}_k s_{q,k} \big) \bm U_T^H \bm \theta_q + \bm w_q \\
	& = \bm U_R \widetilde{\bm H} \bm \phi_q  + \bm w_q,
\end{aligned}
\end{equation}
where $\widetilde{\bm H}=[\widetilde{\bm H}_1,\ldots,\widetilde{\bm H}_K]\in \mathbb{C}^{G_r\times K G_t}$ and $\bm \phi_q \triangleq (\bm s_q \otimes \bm I) \bm U_T^H \bm \theta_q \in \mathbb{C}^{K G_t\times 1}$ with $\bm s_q=[s_{q,1},\cdots,s_{q,K}]^T\in \mathbb{C}^{K \times 1}$. Collecting the received signals over $Q$ time slots  gives
\begin{equation} \label{eq:Y}
		\bm Y  = [\bm y_1,\ldots,\bm y_Q ]= \bm U_R \widetilde{\bm H} \bm \Phi + \bm W \in \mathbb{C}^{M \times Q}
\end{equation}
where $\bm \Phi = [\bm \phi_1,\ldots,\bm \phi_Q]\in \mathbb{C}^{K G_t\times Q}$ and $\bm W = [\bm w_1,\ldots,\bm w_Q]\in \mathbb{C}^{M\times Q}$. Then, the one-bit quantized received signal is given by
\begin{equation} \label{eq:R}
		\bm R = \text{sgn} (\bm Y) \in \mathcal{S}^{M\times Q}.
\end{equation}
Denote $\bm y = \text{Vec}(\bm Y)\in \mathbb{C}^{QM\times 1}$, $\bm h = \text{Vec}(\widetilde{\bm H})\in \mathbb{C}^{K G_r G_t \times 1}$, $\bm w = \text{Vec}(\bm W)\in \mathbb{C}^{QM\times 1}$ and $\bm r = \text{Vec}(\bm R)$. The Eqns.~\eqref{eq:Y} and~\eqref{eq:R} can be rewritten resp. as
\begin{equation} \label{eq: SMV_model_y}
\bm y = \bm \Xi \bm h + \bm w \in \mathbb{C}^{QM\times 1},
\end{equation}
\begin{equation} \label{eq: SMV_model_one_bit}
\bm r =\text{sgn} (\bm y)= \text{sgn} (\bm \Xi \bm h + \bm w)\in \mathcal{S}^{QM\times 1},
\end{equation}
where $\bm \Xi = \bm \Phi^T \otimes \bm U_R \in \mathbb{C}^{QM\times K G_r G_t}$. Since $\bm w\sim \mathcal{CN}(\bm 0,\sigma^2 \bm I )$, it holds that
\begin{equation}\label{eq:y_given_h}
	p(\bm y|\bm h) = \frac{1}{(\pi \sigma^2)^{QM}} \exp\left\{-\frac{\|\bm y - \bm \Xi \bm h\|^2}{\sigma^2} \right\}
\end{equation}

With~\eqref{eq: SMV_model_one_bit} and the VAD representation in~\eqref{eq:VAD_rep},  the cascaded channel estimation problem boils down to estimating $\bm h$ from the one-bit quantized measurements $\bm r$.  Clearly, after estimating $\bm h$,  the path gains $\{\zeta_{\ell_1}^G, \zeta_{\ell_2,k}^r\}_{\ell_1,
		\ell_2}$ and the AoD/AoAs in~\eqref{eq:cascaded_ch} can be obtained from the non-zero coefficients of $\bm h$ and its corresponding support set, resp. Before delving into the details of our proposed method, we should briefly	discuss a related algorithm EM-BPDN in~\cite{stockle2016channel}, which will shed light on our method.


%

%


\subsection{EM-BPDN~\cite{stockle2016channel} } \label{sec:EM-BPDN}
The main idea of EM-BPDN  is to treat $\bm h$ as the unknown parameter and estimate  $\bm h$ via  maximum a posteriori (MAP):
\begin{equation}\label{eq:em_bpdn_main}
\max_{\bm h}~ \ln p(\bm h|\bm r).
\end{equation}
Due to the one-bit quantization, the posterior $p(\bm h|\bm r)$ is hard to compute and generally has no analytical form. To circumvent this difficulty, the work~\cite{stockle2016channel} employs EM method to iteratively solve problem~\eqref{eq:em_bpdn_main}. Specifically, by treating the unquantized data $\bm y$ as a hidden variable, the EM algorithm repeatedly performs the following two steps
\begin{subequations} \label{eq:EM}
\begin{align}
& \textsf{E-step:}~~ \mathcal{Q} (\bm h,\hat{\bm h}^{(\ell)}) \triangleq  \mathbb{E}_{p(\bm y|\bm r,\hat{\bm h}^{(\ell)})}\left[\ln \big(p (\bm r,\bm y,\bm h) \big) \right], \label{eq:EM_E}\\
& \textsf{M-step:}~~ \hat{\bm h}^{(\ell+1)}=\arg\max_{\bm h} \mathcal{Q} (\bm h,\hat{\bm h}^{(\ell)}), \label{eq:EM_M}
\end{align}
\end{subequations}
for every iteration $\ell$ until some stopping criterion is satisfied. Notice that
\[ p (\bm r,\bm y,\bm h)= p (\bm r|\bm y,\bm h) p (\bm y,\bm h) = p (\bm r|\bm y) p (\bm y,\bm h) \]
where the second equality is due to Markov chain $\bm h \rightarrow \bm y \rightarrow \bm r$. Since $p (\bm r|\bm y) $ is independent of $\bm h$, the \textsf{M-step} in \eqref{eq:EM_M}  can be simplified as
\begin{equation} \label{eq:M_step2}
\begin{aligned}
\hat{\bm h}^{(\ell+1)} & = \arg \max_{\bm h} \mathbb{E}_{p(\bm y|\bm r,\hat{\bm h}^{(\ell)})}\left[\ln \big(p (\bm y|\bm h)  \big) + \ln \big(p (\bm h)  \big)  \right] \\
&= \arg \min_{\bm h} \|\bm \Xi \bm h-\hat{\bm y}^{(\ell)} \|^2- \sigma^2 \ln \big(p(\bm h) \big),
\end{aligned}
\end{equation}
where $\hat{\bm y}^{(\ell)}  =  \mathbb{E}_{p(\bm y|\bm r,\hat{\bm h}^{(\ell)})}[\bm y]$ is the posterior mean; $p(\bm h)$ is a pre-specified prior distribution on $\bm h$. In~\cite{stockle2016channel},   the Laplace prior
\begin{equation} \label{eq: lap_prior}
p (\bm h) = \left({\eta^2}/{2 \pi} \right)^{K G_r G_t} \exp (- \eta \|\bm h \|_1)
\end{equation}
with the parameter $\eta>0$ is assumed  to promote  solution sparsity. With~\eqref{eq: lap_prior}, problem~\eqref{eq:M_step2} is a standard basis pursuit denoise (BPND) problem~\cite{Van2008probing}, and is solved by the fast iterative shrinkage-thresholding algorithm (FISTA)~\cite{beck2009fast} in~\cite{stockle2016channel}.

Despite the simplicity of EM-BPDN, there are some drawbacks or limitations when it is used for IRS channel estimation:
\begin{enumerate}
\item The preformance of EM-BPDN relies heavily on the parameter $\eta$, which should be judiciously selected according to the problem instance. Otherwise, the solution of EM-BPDN may suffer from structural error~\cite{wipf2004sparse} and leads to severe performance degradation.
\item  Due to modeling limitation, the EM-BPDN, which was originally proposed for MIMO channels, cannot capture more structured sparsity possessed by the IRS channels. However, as we will see, the latter is the crux of developing an accurate estimator for the IRS channels.
\end{enumerate}

To avoid the above limitations, in the ensuring sections, we will employ more powerful hierarchical SBL  with the variational EM approach to custom-derive  one-bit estimators for the IRS channels.

\section{SBL-Based One-Bit Channel Estimation} \label{sec:SBL}
This section describes the proposed SBL estimator. In the first subsection, we will give a high-level description of the SBL estimator, and the detailed algorithmic derivations are given in the  second and the third subsections.
\subsection{A High-level Description of the SBL Estimator}
Instead of modeling $\bm h$ as the unknown parameter and  estimating it directly,  the  SBL estimator treats both $\bm  h$ and $\bm y$ as  hidden variables, and adopts a  hierarchical SBL model  to  estimate $\bm  h$ indirectly. Specifically, the key idea is to specify a {\it hyperparametric} prior model $p(\bm h; \bm \alpha)$ for the hidden variable $\bm h$ with $\bm \alpha$ being the hyperparameter [cf.~\eqref{eq:prior_dis}]. Under this hierarchical learning model, we first estimate  the hyperparameter $\bm \alpha$ via the following type-II ML:
\begin{equation} \label{eq:var_em}
\max_{\bm \alpha} ~ \ln p(\bm r; \bm \alpha) \triangleq \ln \left( \int\hspace{-5pt} \int p(\bm r, \bm y, \bm h; \bm \alpha ) d \bm y d \bm h \right)
\end{equation}
and then, infer  the hidden variable $\bm h$ as its posterior mean
$$\hat{ \bm h} = \mathbb{E}_{p(\bm h|\bm r; \hat{ \bm \alpha})}[\bm h].$$

Clearly, the advantage of the SBL estimator is that all the variables and parameters are estimated from the data and no parameter like $\eta$ in EM-BPDN needs to be fine-tuned. But the downside is that the  EM method is not feasible for problem~\eqref{eq:var_em}, because  when both $\bm y$ and $\bm h$ are hidden variables,  it is challenging  to calculate the expectation of the posterior distribution $p(\bm y, \bm h| \bm r, \hat{\bm {\alpha}}^{(\ell)})$ in the \textsf{E-step}. To circumvent this difficulty,  a more powerful  variational EM approach is employed to solve~\eqref{eq:var_em}. The key idea of variational EM is to  use  the mean-field variational inference~\cite{hoffman2013stochastic}  to    approximate
the difficult posterior distribution $p(\bm y, \bm h| \bm r, {\bm {\alpha}})$, and employ the  block-coordinate ascent method together with the \textsf{M-step} to iteratively refine the approximated  posterior as well as the hyperparameter $\bm \alpha$. 

%

\subsection{The SBL Approach}

The hyperparametric prior model $p(\bm h; \bm \alpha)$ provides the flexibility to capture the structured sparsity of the IRS channels. Herein, let us first consider the basic element-wise sparsity as in EM-BPDN, and more complex structured sparsity will be discussed in Sec.~\ref{sec:BSBL} and \ref{sec:two_stage}. To this end, the following parameterized zero-mean Gaussian model is  assumed:
\begin{equation} \label{eq:prior_dis}
	p(\bm h; \bm \alpha ) = \mathcal{CN}(\bm h|\bm 0, \bm A) = \prod_{n=1}^{K G_r G_t} (\pi \alpha_{n})^{-1} \exp\left(- \frac{|h_{n}|^2}{\alpha_{n}} \right),
\end{equation}
where $\bm A= \text{Diag}(\bm \alpha)$; $\bm \alpha = [\alpha_1,\ldots,\alpha_{K G_r G_t} ]^T$ is the hyperparameter controlling the prior variance of each entry of $\bm h$. 
We should mention that the  diagonal covariance matrix $\bm A$ can well capture the statistical characteristic of the sparse channel vector $\bm h$, because one can easily show that for independent $\zeta_{\ell_1}^G$ and $\zeta_{\ell_2,k}^r$ with zero mean, each entry of $\bm h$ is uncorrelated.
From~\eqref{eq:prior_dis}, it is clear that the $n$-th entry $ |h_{n}|\rightarrow 0$ as the hyperparameter $\alpha_n\rightarrow 0$~\cite{wipf2004sparse}. Hence, the sparse channel estimation for  $\bm h$ reduces to estimating the hyperparameter $\bm \alpha$. Once $\bm \alpha$ is determined, the $\bm h$ can be set as its mean of the posterior.

The hyperparameter $\bm \alpha$ can be determined by the type-II ML estimation, also known as the evidence maximization, which
marginalizes over the unquantized received signal $\bm y$ and the channel vector $\bm h$, and then performs marginal likelihood maximization. The marginal likelihood is given by
\begin{equation} \label{eq:evidence}
\begin{aligned}
	p(\bm r; \bm \alpha) & = \int\hspace{-5pt} \int p(\bm r, \bm y, \bm h; \bm \alpha ) d \bm y d \bm h \\
	 & =\hspace{-5pt}  \int\hspace{-5pt} \int p(\bm r|\bm y) p(\bm y|\bm h)p(\bm h; \bm \alpha )d \bm y d \bm h \\
	 & =\hspace{-5pt} \int\hspace{-5pt} \int \mathbb{I}_{\text{sgn}(\bm y)}(\bm r) \mathcal{CN}(\bm y| \bm \Xi \bm h, \sigma^2 \bm I) \mathcal{CN} (\bm h|\bm 0, \bm A ) d \bm y d \bm h,
\end{aligned}
\end{equation}
where $\mathbb{I}_{\text{sgn}(\bm y)}(\bm r)$ is an element-wise indicator function
\begin{numcases}{\mathbb{I}_{\text{sgn}(y)}( r)=}
1, ~~~r= \text{sgn}(y), \notag \\
0, ~~~\text{otherwise}.
\end{numcases}
Directly maximizing $p(\bm r; \bm \alpha) $  is intractable. A more practical way
is to use the EM method to iteratively maximize $p(\bm r; \bm \alpha)$. By treating $\bm y$ and $\bm h$ as hidden variables, the EM repeatedly   performs the following two steps:
\begin{itemize}
	\item  \textsf{E-step}: Compute the expected log-likelihood of the complete data set $\{\bm r, \bm y,\bm h\}$:
	\begin{equation*}
	\begin{aligned}
		\mathcal{Q} (\bm \alpha,\hat{\bm \alpha}^{(\ell)})& \triangleq \mathbb{E}_{p(\bm y, \bm h|\bm r;\hat{\bm \alpha}^{(\ell)} )}\left[\ln p(\bm r, \bm y, \bm h; \bm \alpha) \right] \\
		& = \mathbb{E}_{p(\bm y, \bm h|\bm r;\hat{\bm \alpha}^{(\ell)} )} \big[\ln p(\bm r|\bm y) p(\bm y|\bm h) \big] \\
		& \hspace{10pt} + \mathbb{E}_{p(\bm y, \bm h|\bm r;\hat{\bm \alpha}^{(\ell)} )} \big[\ln p(\bm h; \bm \alpha ) \big],
	\end{aligned}
	\end{equation*} 	
	\item \textsf{M-step}: Revise the hyperparameter estimate $\hat{\bm \alpha}^{(\ell+1)}$ by maximizing $\mathcal{Q} (\bm \alpha,\hat{\bm \alpha}^{(\ell)})$:
	\begin{subequations} \label{eq:Mstep_alpha}
		\begin{align}
		&\hat{\bm \alpha}^{(\ell+1)}  = \arg\max_{\bm \alpha \in \mathbb{R}_{+}^{K G_r G_t \times 1}} \mathcal{Q} (\bm \alpha,\hat{\bm \alpha}^{(\ell)})\label{eq:Mstep_alpha_a} \\
		& = \arg\max_{\bm \alpha \in \mathbb{R}_{+}^{K G_r G_t \times 1}} \mathbb{E}_{p(\bm y, \bm h|\bm r;\hat{\bm \alpha}^{(\ell)} )}\left[\ln p(\bm h;\bm \alpha) \right],	\label{eq:Mstep_alpha_b}		
		\end{align}
	\end{subequations}
	where~\eqref{eq:Mstep_alpha_b} follows from the fact that $\mathbb{E}_{p(\bm y, \bm h|\bm r;\hat{\bm \alpha}^{(\ell)})} \big[\ln p(\bm r|\bm y) p(\bm y|\bm h) \big] $ does not depend on $\bm \alpha$.
	\end{itemize}
By substituting~\eqref{eq:prior_dis} into~\eqref{eq:Mstep_alpha_b}, the $n$-th element of $\hat{\bm \alpha}^{(\ell+1)}$ is given by
\begin{subequations} \label{eq:Mstep_alpha2}
	\begin{align}
	\hat{\alpha}_{n}^{(\ell+1)} &=\arg\min_{\alpha_{n} \in \mathbb{R}_{+}} \ln \alpha_{n} + \frac{\mathbb{E}_{p(\bm y, \bm h|\bm r;\hat{\bm \alpha}^{(\ell)})}[|h_{n}|^2]}{\alpha_{n}} \label{eq:Mstep_alpha2_a} \\
	&=\mathbb{E}_{p(\bm y, \bm h|\bm r;\hat{\bm \alpha}^{(\ell)})}[|h_{n}|^2]	  \label{eq:Mstep_alpha2_b}		
	\end{align}
\end{subequations}
for $n=1,\ldots, K G_r G_t$.

Clearly, it suffices to compute $E_{p(\bm y, \bm h|\bm r;\hat{\bm \alpha}^{(\ell)})}[|h_{n}|^2]$ to run the EM algorithm. However, due to  the one-bit quantization,  it is challenging to evaluate the posterior $p(\bm y, \bm h| \bm r; \hat{\bm \alpha}^{(\ell)})$ or compute its expectation. In the next subsection, we leverage on the variational EM  to approximately
evaluate the posterior distribution $p(\bm y, \bm h| \bm r; \hat{\bm \alpha}^{(\ell)})$ so that the expectation in~\eqref{eq:Mstep_alpha2_b} can be efficiently evaluated.


\subsection{A Variational EM Approach to~\eqref{eq:Mstep_alpha2_b}} \label{sec: V-EM}
The key idea of the variational EM is to approximate the complex distribution $p(\bm y, \bm h| \bm r; \bm \alpha)$ with a simpler one $q(\bm y, \bm h)$, which  takes the following factorized form
\begin{equation} \label{eq:q}
p(\bm y, \bm h| \bm r; \bm \alpha) \approx q(\bm y, \bm h) \triangleq q(\bm y) q(\bm h),
\end{equation}
so that the expectation in~\eqref{eq:Mstep_alpha2_b} can be efficiently evaluated (with closed form) w.r.t.  $q(\bm y)$ for fixed $q(\bm h)$, and vice versa. In essence, the variational EM can be seen as an approximate EM with the \textsf{M-step} completed by block-coordinate ascent (BCA). To find a good approximation $q(\bm y, \bm h)$,
the minimum Kullback-Leibler (KL) divergence between $q(\bm y, \bm h)$ and $p(\bm y, \bm h| \bm r; \bm \alpha)$ is sought, viz.,
\begin{equation} \label{eq:KL_min}
	\begin{aligned}
		\min_{q(\bm y,\bm h)}& ~~ -\mathbb{E}_{q(\bm y,\bm h)} \left\{\ln \frac{p(\bm y, \bm h| \bm r; \bm \alpha)}{q(\bm y,\bm h)} \right\} \\
		\text{s.t.} & ~ ~q(\bm y, \bm h) = q(\bm y) q(\bm h).
	\end{aligned}
\end{equation}
Note that while problem~\eqref{eq:KL_min} is known to be non-convex due to the non-convexity of the mean-field family set, it is convex w.r.t  $q(\bm y)$ for fixed  $q(\bm h)$, and vice versa. Therefore,  $q(\bm y)$ and $q(\bm h)$ can be alternately optimized.

\subsubsection{ Optimizing $q(\bm y)$ for fixed $q(\bm h)$}
According to~\cite{tzikas2008variational}, for fixed $q(\bm h)$ the optimal $q^\star(\bm y)$ in~\eqref{eq:KL_min} should satisfy the following relation:
\begin{equation} \label{eq:ln_q_yR}
	\ln q^{\star}(\bm y) = \mathbb{E}_{q(\bm h)} [\ln p(\bm r, \bm y, \bm h; \bm \alpha)] + \text{const},
\end{equation}
where  the  constant  in~\eqref{eq:ln_q_yR} can be obtained via normalization (if required). Using the decomposition $p(\bm r, \bm y, \bm h; \bm \alpha) = p(\bm r| \bm y)p(\bm y| \bm h) p(\bm h; \bm \alpha)$ and substituting~\eqref{eq:y_given_h} for the conditional distribution $p(\bm y| \bm h)$, we have
\begin{equation} \label{eq:ln_q_yR2}
	\begin{aligned}
\ln q^{\star}(\bm y) &=\ln p(\bm r| \bm y)+\mathbb{E}_{q(\bm h)} [\ln p(\bm y| \bm h)]+ \text{const}\\
		& = \sum_{i=1}^{Q M} \ln p(r_{i}|y_{i})-\sum_{i=1}^{Q M}\frac{|y_{i} - \bm \xi_{i}^T \mathbb{E}_{q(\bm h)}[\bm h] |^2}{\sigma^2}+ \text{const}
	\end{aligned}
\end{equation}
where $\bm \xi_{i}^T$ is the $i$-th row of $\bm \Xi$ and the terms irrespective of $\bm y$ are absorbed into the constant.


\subsubsection{Optimizing $q(\bm h)$ for fixed $q(\bm y)$}
Similarly, for fixed $q(\bm y)$  the optimal $q^\star(\bm h)$ is given by
 \begin{equation}\label{eq:ln_qh2}
 	\begin{aligned}
 		\ln q^{\star}(\bm h)  = &-\bm h^H\left(\frac{ \bm \Xi^H \bm \Xi}{\sigma^2}\right)\bm h- \bm h^H \text{Diag}(\bm \alpha)^{-1} \bm h \\
 		& +  \frac{(\mathbb{E}_{q(\bm y)}[\bm y])^H \bm \Xi \bm h +\bm h^H \bm \Xi^H \mathbb{E}_{q(\bm y)}[\bm y]}{\sigma^2} + \text{const}.
 	\end{aligned}
 \end{equation}
Since the right-hand side of~\eqref{eq:ln_qh2} is a quadratic function of $\bm h$, the distribution $q^{\star}(\bm h)$ can be recognized as Gaussian distribution. Hence, we can complete the square over $\bm h$ to identify the mean and covariance:
\begin{subequations} \label{eq:mu_Sigma}
 \begin{align}
  \bm \mu_h & \triangleq \mathbb{E}_{q(\bm h)}[\bm h] = \sigma^{-2} \bm \Sigma_h \bm \Xi^H \mathbb{E}_{q(\bm y)}[\bm y], \label{eq:mu_Sigma_a} \\
  \bm \Sigma_h & = \left({\text{Diag}(\bm \alpha)}^{-1} +\sigma^{-2} \bm \Xi^H \bm \Xi \right)^{-1}. \label{eq:mu_Sigma_b}
 \end{align}
\end{subequations}
Then $q^{\star}(\bm h)$ is given by
 \begin{equation}\label{eq:optimal_q_h_R}
 	q^{\star}(\bm h) = \mathcal{CN}(\bm h | \bm \mu_h, \bm \Sigma_h).
 \end{equation}
Moreover, the mean vector $\bm \mu_y \triangleq \mathbb{E}_{q(\bm y)}[\bm y]$ in~\eqref{eq:mu_Sigma_a} can be computed in an element-wise manner, i.e., (see the supplementary material for the derivation)
\begin{equation} \label{eq:mu_y}
	\begin{aligned}
		\mu_{y,i} &=\mathbb{E}_{q(y_{i})}[y_i] \\
		 & = \frac{\sigma}{\sqrt{2}}\left(\mathfrak{R}\{r_i\} \frac{\psi ( \chi_i^R)}{\Psi(  \chi_i^R )} +j\cdot \mathfrak{I}\{r_i\}\frac{\psi ( \chi_i^I)}{\Psi(  \chi_i^I )} \right) +z_i
	\end{aligned}
\end{equation}
for $i=1,\ldots,Q M$, where $\mu_{y,i}$ is the $i$-th element of $\bm \mu_y$, $z_i = \bm \xi_{i}^T \mathbb{E}_{q(\bm h)}[\bm h]$, $\chi^R_i = \mathfrak{R}\{r_i\}\mathfrak{R}\{z_i\}  /\sqrt{\sigma^2/2}$, $\chi^I_i = \mathfrak{I}\{r_i\}\mathfrak{I}\{z_i\}  /\sqrt{\sigma^2/2}$, $\psi(x)=\frac{1}{\sqrt{2\pi}} \exp(-\frac{x^2}{2})$ and $\Psi(x) = \int_{-\infty}^{x} \psi (t) dt$.


\subsubsection{Evaluating~\eqref{eq:Mstep_alpha2_b} for fixed $q(\bm h)$ and $q(\bm y)$}
Now, we are ready to approximately compute the expectation $\mathbb{E}_{p(\bm y, \bm h|\bm r;\hat{\bm \alpha}^{(\ell)})}[|h_{n}|^2]$ in~\eqref{eq:Mstep_alpha2_b}. Let $q^{(\ell)}(\bm y,\bm h)=q^{(\ell)}(\bm y)q^{(\ell)}(\bm h)$ be the variational approximation of the posterior $p(\bm y, \bm h| \bm r; \hat{\bm \alpha}^{(\ell)})$ at the $\ell$-th EM iteration, we have
\begin{subequations} \label{eq:appro_Eh}
	\begin{align}
	& \mathbb{E}_{p(\bm y, \bm h|\bm r;\hat{\bm \alpha}^{(\ell)})}[|h_{n}|^2] \\
	= &  \int  \int |h_{n}|^2 p(\bm y, \bm h| \bm r; \hat{\bm \alpha}^{(\ell)}) d \bm y d \bm h \label{eq:appro_Eh_a} \\
	\approx &  \int  \int |h_{n}|^2 q^{(\ell)}(\bm y) q^{(\ell)}(\bm h) d \bm y d \bm h \label{eq:appro_Eh_b} \\
	= & \int |h_{n}|^2 q^{(\ell)}(\bm h)  d \bm h \label{eq:appro_Eh_c} \\
	= & |[\bm \mu_h^{(\ell)}]_n|^2 + [\bm \Sigma_h^{(\ell)}]_{n,n} \label{eq:appro_Eh_d}
	\end{align}
\end{subequations}
for $n=1,\ldots,K G_r G_t$, where ~\eqref{eq:appro_Eh_b} is due to $p(\bm y, \bm h| \bm r; \hat{\bm \alpha}^{(\ell)}) \approx q^{(\ell)}(\bm y, \bm h) = q^{(\ell)}(\bm y) q^{(\ell)}(\bm h)$; $[\bm \mu_h^{(\ell)}]_n$ and $[\bm \Sigma_h^{(\ell)}]_{n,n}$ are the $n$-th and $(n,n)$-th elements of the a posteriori mean vector $\bm \mu_h^{(\ell)}$ and covariance matrix $\bm \Sigma_h^{(\ell)}$ for the variational posterior $q^{(\ell)}(\bm h)$, which are calculated via~\eqref{eq:mu_Sigma} with $\bm \alpha$ in $\bm A = \text{Diag}(\bm \alpha)$ replaced by $\hat{\bm \alpha}^{(\ell)}$. Therefore, the $n$-th element of the new estimate $\hat{\bm \alpha}^{(\ell+1)}$ in~\eqref{eq:Mstep_alpha2} is approximately updated as
\begin{equation} \label{eq:Mstep_alpha3}
	\hat{\alpha}_n^{(\ell+1)}=\mathbb{E}_{p(\bm y, \bm h|\bm r;\hat{\bm \alpha}^{(\ell)})}[h_{n}^2]\approx |[\bm \mu_h^{(\ell)}]_n|^2 + [\bm \Sigma_h^{(\ell)}]_{n,n}
\end{equation}
for $n=1,\ldots, K G_r G_t$.

Algorithm~\ref{alg:2} summarizes the whole procedure of SBL-based one-bit channel estimation via variational EM. Note that the matrix inversion in line 6 may be cheaply computed by applying the  matrix inversion lemma:
\begin{equation}
	\begin{aligned}
\bm \Sigma_h^{(\ell)} &=\left({\text{Diag}(\hat{\bm \alpha}^{(\ell)})}^{-1} +\sigma^{-2} \bm \Xi^H \bm \Xi \right)^{-1} \\
&=\bm A^{(\ell)} - \bm A^{(\ell)} \bm \Xi^H \left(\sigma^{2}\bm I  +\bm \Xi \bm A^{(\ell)} \bm \Xi^H\right)^{-1} \bm \Xi \bm A^{(\ell)} 		
	\end{aligned}
\end{equation}
where $\bm A^{(\ell)}  = \text{Diag}(\hat{\bm \alpha}^{(\ell)})$.
In addition, from line 8 to line 11  we have an inner cyclical update of $\bm \mu_h^{(\ell)}$, because given the hyperparameter $ \hat{\bm \alpha}^{(\ell)}$, the evaluations of $q^{(\ell)}(\bm y)$ and $q^{(\ell)}(\bm h)$ are coupled---in \eqref{eq:ln_q_yR2} $q^\star(\bm y)$ depends on $\mathbb{E}_{q(\bm h)}[\bm h]$ and  in \eqref{eq:optimal_q_h_R} $q^\star(\bm h)$ depends on $\mathbb{E}_{q(\bm y)}[\bm y]$. Hence, it is needed to alternately optimize $q^{(\ell)}(\bm y)$ and $q^{(\ell)}(\bm h)$ until convergence before running  the $\textsf{M-step}$ in line 14. Nevertheless, by our numerical experience, the inner iteration number $p_{\max}$ is usually small, typically less than ten iterations. After convergence,  the channel vector $\bm h$ is inferred as the mean of  $q^{(\ell)}(\bm h)$, i.e., $\bm h=\bm \mu_{h}^{(\ell)}$.

\begin{algorithm}
	\caption{SBL-based One-bit Channel Estimation} \label{alg:2}
	\begin{algorithmic}[1]
		\State Input $\bm \Xi$, $\bm r$, $\sigma^2$, $\hat{\bm \alpha}^{(0)}=0.001\times \bm 1$ and $p_{\text{max}}$.
		\State Set $\ell=-1$ and $\bm \mu_h^{(-1)}= \bm \Xi^{\dag} \bm r$		
		\Repeat
		\State $\ell \leftarrow \ell+1$
		\State {\bf Variational E-step}: given $\hat{\bm \alpha}^{(\ell)}$ compute
		\State $\bm \Sigma_h^{(\ell)} =\left({\text{Diag}(\hat{\bm \alpha}^{(\ell)})}^{-1} +\sigma^{-2} \bm \Xi^H \bm \Xi \right)^{-1}$
		\State Set $p=0$ and $\bm \mu_h^{(\ell-1)(0)}= \bm \mu_h^{(\ell-1)}$,	
		 \Repeat	
		 \State $p \leftarrow p+1$
		 \begin{align*}
		 	\mu_{y,i}^{(p)} & =\frac{\sigma}{\sqrt{2}}\left(\mathfrak{R}\{r_i\} \frac{\psi ( \chi_i^R)}{\Psi(  \chi_i^R )} +j\cdot \mathfrak{I}\{r_i\}\frac{\psi ( \chi_i^I)}{\Psi(  \chi_i^I )} \right) \\
		 	& +\bm \xi_{i}^T \bm \mu_h^{(\ell-1)(p-1)},~ \forall i
		 \end{align*}
		 \State $\bm \mu_h^{(\ell-1)(p)}   =\sigma^{-2} \bm \Sigma_h^{(\ell)} \bm \Xi^H \bm \mu_{y}^{(p)}	$
		 \Until{$p>p_{\text{max}}$}
		 \State $\bm \mu_h^{(\ell)}=\bm \mu_h^{(\ell-1)(p)}$.
		 \State {\bf Variational M-step}: update the hyperparemeters
		 \State $\hat{\alpha}_n^{(\ell+1)}= |[\bm \mu_h^{(\ell)}]_n|^2 + [\bm \Sigma_h^{(\ell)}]_{n,n},~ \forall n=1,\ldots,K G_r G_t.$
		\Until{some stopping criterion is satisfied}
		\State Output: ${\bm h}=\bm \mu_h^{(\ell)}$.
		\State Recover ${\widetilde{\bm H}} = \text{Vec}^{-1} ({\bm h})$ from ${\bm h}$, the cascaded channels can be estimated by ${\bm H}_k = \bm U_R {\widetilde{\bm H}}_k \bm U_T^H$, where ${\widetilde{\bm H}}_k = [\widetilde{\bm H}]_{:, (k-1)G_t+1:kG_t}$, $\forall k$.
	\end{algorithmic}
\end{algorithm}

\section{Block SBL-based One-Bit Channel Estimation}\label{sec:BSBL}
\begin{figure}[!h]
	\centerline{\resizebox{.25\textwidth}{!}{\includegraphics{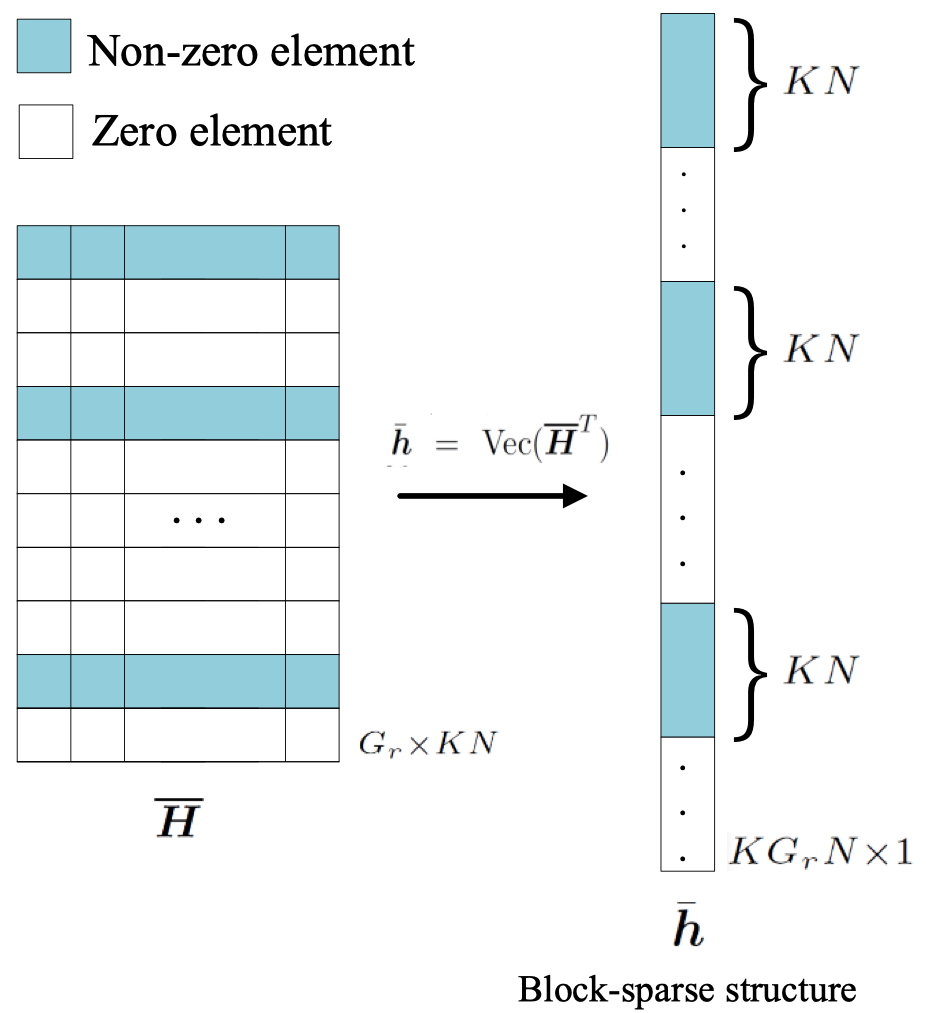}}
	}   \caption{The block-sparse structure of $\bar{\bm h} $.} \label{fig:block_sparse_structure}
\vspace{-10pt}
\end{figure}

The SBL estimator utilizes the elementwise sparsity of $\{\widetilde{\bm H}_k\}_{k\in \cal K}$. However, since all the users communicate with the BS via the IRS, the $\{\widetilde{\bm H}_k\}_{k\in \cal K}$  share some common sparse structure induced by the IRS-to-BS channel; see Fig.~\ref{fig:block_sparse_structure} for an illustration. In  this section, we will take advantage of the common row-structured sparsity to improve the SBL performance.

Let us first formulate the channel estimation as a multiple measurement vector (MMV) recovery problem.
Denote $\overline{\bm H}_k \triangleq \widetilde{\bm H}_k \bm U_T^H$, $\forall k$. The signal model in~\eqref{eq:y_q2} can be rewritten as
\begin{equation} \label{eq:y_q_BSBL}
\begin{aligned}
	\bm y_q & = \bm U_R \big(\sum_{k=1}^{K} \overline{\bm H}_k s_{q,k} \big) \bm \theta_q + \bm w_q \\
	& = \bm U_R \overline{\bm H} \bm \triangle_q + \bm w_q
\end{aligned}
\end{equation}
for $q=1,\ldots,Q$, where $\overline{\bm H} = [ \overline{\bm H}_1,\ldots,  \overline{\bm H}_K]\in \mathbb{C}^{G_r\times KN}$ and $\bm \triangle_q = (\bm s_q \otimes \bm I)\bm \theta_q \in \mathbb{C}^{KN\times 1}$. By defining $\bm Y=[\bm y_1,\ldots,\bm y_Q ]\in \mathbb{C}^{M\times Q}$ as a measurement matrix, one can express the received signals as the MMV model
\begin{equation} \label{eq:MMV_model}
	\bm Y = \bm U_R \overline{\bm H} \bm \Delta + \bm W,
\end{equation}
where $\bm \Delta = [\bm \triangle_1,\ldots,\bm \triangle_Q]\in \mathbb{C}^{KN\times Q}$. For the IRS-aided communication systems, the channel $\bm G$ from the IRS to the BS is common for all the users; this can also be seen from~\eqref{eq:cascaded_ch}, where the AoA parameters $\{\vartheta_{\ell_1}^{AoA}\}$ are independent of the user index $k$. Hence, each column of the concatenated channel matrix $\overline{\bm H}$ should  share the same sparsity profile, as depicted in Fig~\ref{fig:block_sparse_structure}.

By letting $\bar{\bm y}=\text{Vec}(\bm Y^T)\in \mathbb{C}^{MQ\times 1}$, $\bm \Upsilon=\bm U_R \otimes \bm \Delta^T \in \mathbb{C}^{QM\times K G_r N}$, $\bar{\bm h}=\text{Vec}(\overline{\bm H}^T)\in \mathbb{C}^{K G_r N\times 1}$ and $\bar{\bm w}=\text{Vec}(\bm W^T)\in \mathbb{C}^{MQ\times 1}$, the MMV model is recast into the single measurement vector (SMV) model
\begin{equation} \label{eq:block_SMV_model}
	\bar{\bm y} =\bm \Upsilon \bar{\bm h} + \bar{\bm w}
\end{equation}
with block sparse pattern. The corresponding one-bit quantized measurement is given by
\begin{equation} \label{eq:block_SMV_onebit_model}
	\bar{\bm r} = \text{sgn}({ \bar{\bm y}}) =\text{sgn}(\bm \Upsilon \bar{\bm h} + \bar{\bm w}).
\end{equation}

To exploit the block sparsity in $\bar{\bm h}$,  we consider the following block SBL (BSBL)-based prior model on $\bar{\bm h}$:
\begin{equation} \label{eq:block_prior_model}
  p(\bar{\bm h};\bm \gamma,\bm B)=\prod_{n=1}^{G_r} p(\bar{\bm h}^{[n]};\gamma_{n},\bm B)=\mathcal{CN}(\bm 0, \bm \Sigma_0),
\end{equation}
where $\bar{\bm h}^{[n]}\triangleq [\bar{\bm h}]_{(n-1)KN+1: nKN}$ denotes the $n$-th block of $\bar{\bm h}$; $\bm \gamma =[\gamma_{1},\ldots,\gamma_{G_r}]^T \in \mathbb{R}_+^{G_r\times 1}$  and  $\bm B \in \mathbb{H}_{++}^{KN}$ are hyperparameters, which capture the block sparsity of $\bar{\bm h}$ and the intra-block correlation through the covariance matrix ${\bm \Sigma_0}$:
\[ {\bm \Sigma_0} = {\rm Diag}( \gamma_{1}\bm B, \ldots, \gamma_{G_r} \bm B)  =\bm \Gamma \otimes \bm B \]
where  $\bm \Gamma = \text{Diag}(\bm \gamma)$.

Similar to the  SBL, the hyperparameters $\bm \gamma$ and $\bm B$ are estimated via maximizing the type-II ML, viz.
\begin{equation} \label{eq:evidence_BSBL}
\max_{\bm \gamma, \bm B}~ \ln \left(	p(\bar{\bm r}; \bm \gamma, \bm B) \right) = \ln\left(  \int\hspace{-5pt} \int p(\bar{\bm r}, \bar{\bm y}, \bar{\bm h}; \bm \gamma, \bm B ) d \bar{\bm y} d \bar{\bm h} \right).
\end{equation}
As before,  the variational EM is employed to tackle problem~\eqref{eq:evidence_BSBL}. Let $\{\hat{\bm \gamma}^{(\ell)}, \hat{\bm B}^{(\ell)}\}$ denote the estimate of the hyperparameters in the $\ell$-th EM iteration. The variational EM repeatedly performs the following update:
\begin{equation}  \label{eq:v-em-bsbl}
(\hat{\bm \gamma}^{(\ell+1)}, \hat{\bm B}^{(\ell+1)} ) = \arg\max_{\bm \gamma, \bm B}\mathbb{E}_{q^{(\ell)}(\bar{\bm y},\bar{\bm h})} \left[\ln p(\bar{\bm r}, \bar{\bm y},\bar{\bm h}; \bm \gamma, \bm B) \right]
\end{equation}
where  $q(\bar{\bm y},\bar{\bm h})$ denotes a mean-field variational approximation of the posterior $p(\bar{\bm y},\bar{\bm h}|\bar{\bm r}; {\bm \gamma}, {\bm B})$. For notational simplicity, we will occasionally suppress the iteration index $\ell$, when there is no ambiguity.

Following the same derivation in Section~\ref{sec: V-EM}, the optimal  $q^\star(\bar{\bm h})$ for fixed $q(\bar{\bm y})$ is given by
\begin{equation} \label{eq:bsbl_qh}
 	q^{\star}(\bar{\bm h}) = \mathcal{CN} \left(\bar{\bm h} | \bm \mu_{\bar{h}}, \bm \Sigma_{\bar{h}} \right)
 \end{equation}
  with
 \begin{subequations} \label{eq:mu_Sigma_BSBL}
	\begin{align}
		\bm \mu_{\bar{h}} & \triangleq \mathbb{E}_{q(\bar{\bm h})}[\bar{\bm h}]  = \sigma^{-2} \bm \Sigma_{\bar{h}} \bm \Upsilon^H  \mathbb{E}_{q(\bar{\bm y})}[\bar{\bm y}], \label{eq:mu_Sigma_BSBL_a} \\
		 \bm \Sigma_{\bar{h}} & = \left({\bm \Sigma_0}^{-1} +\sigma^{-2} \bm \Upsilon^H \bm \Upsilon \right)^{-1}. \label{eq:mu_Sigma_BSBL_b}
	\end{align}
\end{subequations}
Similarly, following the same derivation of calculating $\mathbb{E}_{q(\bm y)}[\bm y]$ in the supplementary file, the mean vector of the posterior $q(\bar{\bm y})$, denoted as $\bm \mu_{\bar{y}}\triangleq \mathbb{E}_{q(\bar{\bm y})}[\bar{\bm y}]$, is calculated as
\begin{equation} \label{eq:cal_E_yR_BSBL}
	\begin{aligned}
		\mu_{\bar{y},i} &=\mathbb{E}_{q(\bar{y}_i)}[\bar{y}_{i}] \\
		 & = \frac{\sigma}{\sqrt{2}}\left(\mathfrak{R}\{\bar{r}_i\} \frac{\psi ( \bar{\chi}_i^R)}{\Psi(  \bar{\chi}_i^R )} +j\cdot \mathfrak{I}\{\bar{r}_i\}\frac{\psi (  \bar{\chi}_i^I)}{\Psi(   \bar{\chi}_i^I )} \right) +\bar{z}_i
	\end{aligned}
\end{equation}
for $i=1,\ldots,Q M$, where $\mu_{\bar{y},i}$ is the $i$-th element of $\bm \mu_{\bar{y}}$; $\bar{z}_i=\bm \upsilon_{i}^T \mathbb{E}_{q(\bar{\bm h})}[\bar{\bm h}]$ with $\bm \upsilon_{i}^T$ being the $i$-th row of $\bm \Upsilon$; $\bar{\chi}_i^R = \mathfrak{R}\{\bar{r}_i\}\mathfrak{R}\{\bar{z}_i\}  /\sqrt{\sigma^2/2}$ and $\bar{\chi}_i^I = \mathfrak{I}\{\bar{r}_i\}\mathfrak{I}\{\bar{z}_i\}  /\sqrt{\sigma^2/2}$.

Now, with~\eqref{eq:bsbl_qh}, \eqref{eq:mu_Sigma_BSBL} and \eqref{eq:cal_E_yR_BSBL}, we are ready to calculate the posterior mean in~\eqref{eq:v-em-bsbl}. It can be shown that the \textsf{M-step}  in~\eqref{eq:v-em-bsbl} amounts to
\begin{equation}\label{eq:var_Mstep_d}
(\hat{\bm \gamma}^{(\ell+1)}, \hat{\bm B}^{(\ell+1)}) = \arg\min_{\bm \gamma, \bm B}  \mathcal{F} (\bm \gamma, \bm B ; \hat{\bm \gamma}^{(\ell)} ,\hat{\bm B}^{(\ell)})
\end{equation}
where $\mathcal{F} (\bm \gamma, \bm B,\hat{\bm \gamma}^{(\ell)}, \hat{\bm B}^{(\ell)})  \triangleq KN \ln (|\bm \Gamma|)+G_r \ln(|\bm B|)  + \text{Tr}\big[\big(\bm \Gamma^{-1}\otimes \bm B^{-1}) ( \bm \Sigma_{\bar{h}}^{(\ell)}+\bm \mu_{\bar{h}}^{(\ell)} (\bm \mu_{\bar{h}}^{(\ell)})^H ) \big] $ with $\bm \Sigma_{\bar{h}}^{(\ell)}$ and $\bm \mu_{\bar{h}}^{(\ell)}$ given in~\eqref{eq:mu_Sigma_BSBL}.
Notice that $\gamma_{n}~\forall~n$ are decoupled in~\eqref{eq:var_Mstep_d}. By setting the derivative of $\mathcal{F}$ in~\eqref{eq:var_Mstep_d} w.r.t. $\gamma_{n}$  to zero, we obtain
\begin{equation} \label{eq:gamma_update_BSBL}
	\hat{\gamma}_{n}^{(\ell+1)}=\frac{\text{Tr}\big[(\hat{\bm B}^{(\ell)})^{-1} ( \bm \Sigma_{\bar{h},n}^{(\ell)}+\bm \mu_{\bar{h},n}^{(\ell)} (\bm \mu_{\bar{h},n}^{(\ell)})^H ) \big]}{KN}
\end{equation}
for $n=1,\ldots,G_r$, where the block mean vector $\bm \mu_{\bar{h},n}^{(\ell)}$ and the covariance matrix $\bm \Sigma_{\bar{h},n}^{(\ell)}$ are defined as
\begin{align*}
	&\bm \mu_{\bar{h},n}^{(\ell)} \triangleq [\bm \mu_{\bar{h}}^{(\ell)}]_{(n-1)KN+1:nKN}, \\
	&\bm \Sigma_{\bar{h},n}^{(\ell)} \triangleq  [\bm \Sigma_{\bar{h}}^{(\ell)}]_{(n-1)KN+1 : nKN,(n-1)KN+1 : nKN},
\end{align*}
resp. Similarly, by setting the derivative of $\mathcal{F}$ in~\eqref{eq:var_Mstep_d} w.r.t. $\bm B$ to zero, we obtain
\begin{equation} \label{eq:B_update_BSBL}
	\hat{\bm B}^{(\ell+1)}=\frac{1}{G_r}\sum_{n=1}^{G_r} \frac{\bm \Sigma_{\bar{h},n}^{(\ell)}+\bm \mu_{\bar{h},n}^{(\ell)} (\bm \mu_{\bar{h},n}^{(\ell)})^H}{\hat{\gamma}_{n}^{(\ell+1)}}.
\end{equation}


Algorithm~\ref{alg:3} summarizes the whole procedure of the BSBL estimator. Again, the matrix inversion lemma can be used to alleviate the computational complexity of matrix inversion in line 6. After convergence, the mean of the variational approximation $q^{(\ell)}(\bar{\bm h})$ is used for Bayesian inference for $\bar{\bm h}$, i.e., ${\bar{\bm h}}=\bm \mu_{\bar{h}}^{(\ell)}$. Finally, the  channel for the $k$-th user can be estimated as
\begin{equation}
	{\bm H}_k=\bm U_R {\overline{\bm H}}_k, ~k=1,\ldots,K,	
\end{equation}
where ${\overline{\bm H}}_k = [{\overline{\bm H}}]_{:, (k-1)N+1:kN}$ with ${\overline{\bm H}} = \big(\text{Vec}^{-1}({\bar{\bm h}})\big)^{T}$.
\begin{algorithm}
	\caption{BSBL-based One-bit Channel Estimation} \label{alg:3}
	\begin{algorithmic}[1]
		\State Input $\bm \Upsilon$, $\bar{\bm r}$, $\sigma^2$, $\hat{ \bm \gamma}^{(0)}=0.001\times \bm 1$, $\hat{\bm B}^{(0)}=\bm I$ and $p_{\text{max}}$.
		\State Set $\ell=-1$ and $\bm \mu_{\bar{h}}^{(-1)}= \bm \Upsilon^{\dag} \bar{\bm r}$		
		\Repeat
		\State $\ell \leftarrow \ell+1$
		\State {\bf Variational E-step}: given $\{\hat{\bm \gamma}^{(\ell)}, \hat{\bm B}^{(\ell)}\}$ compute
		\State $\bm \Sigma_{\bar{h}}^{(\ell)} = ((\bm \Sigma^{(\ell)}_0)^{-1}+\sigma^{-1} \bm \Upsilon^H \bm \Upsilon)^{-1}$
		\State Set $p=0$ and $\bm \mu_{\bar{h}}^{(\ell-1)(0)}= \bm \mu_{\bar{h}}^{(\ell -1)}$,	
		 \Repeat	
		 \State $p \leftarrow p+1$
		 \begin{align*}
		 	\mu_{\bar{y},i}^{(p)} & =\frac{\sigma}{\sqrt{2}}\left(\mathfrak{R}\{\bar{r}_i\} \frac{\psi ( \bar{\chi}_i^R)}{\Psi(  \bar{\chi}_i^R )} +j\cdot \mathfrak{I}\{\bar{r}_i\}\frac{\psi (  \bar{\chi}_i^I)}{\Psi(   \bar{\chi}_i^I )} \right)  \\
		 	& +\bm \upsilon_{i}^T \bm \mu_{\bar{h}}^{(\ell -1)(p-1)},~ \forall i
		 \end{align*}		
		 \State $\bm \mu_{\bar{h}}^{(\ell-1)(p)}=\sigma^{-2} \bm \Sigma_{\bar{h}}^{(\ell)} \bm \Upsilon^H \bm \mu_{\bar{y}}^{(p)}$
		 \Until{$p>p_{\text{max}}$}
		 \State $\bm \mu_{\bar{h}}^{(\ell)}=\bm \mu_{\bar{h}}^{(\ell-1)(p)}$.
		 \State {\bf Variational M-step}: update the hyperparemeters
		 \State $\hat{\gamma}_{n}^{(\ell+1)}\hspace{-3pt} =\hspace{-3pt}\text{Tr}\big[(\hat{\bm B}^{(\ell)})^{-1} ( \bm \Sigma_{\bar{h},n}^{(\ell)}+\bm \mu_{\bar{h},n}^{(\ell)} (\bm \mu_{\bar{h},n}^{(\ell)})^H ) \big]/(KN)$
		 \State $\hat{\bm B}^{(\ell+1)} =\frac{1}{G_r}\sum_{n=1}^{G_r} \frac{\bm \Sigma_{\bar{h},n}^{(\ell)}+\bm \mu_{\bar{h},n}^{(\ell)} (\bm \mu_{\bar{h},n}^{(\ell)})^H}{\hat{\gamma}_{n}^{(\ell+1)}}.$
		\Until{some stopping criterion is satisfied}		
		\State Output: ${\bar{\bm h}}=\bm \mu_{\bar{h}}^{(\ell)}$.
		\State Recover ${\overline{\bm H}} = \big(\text{Vec}^{-1}({\bar{\bm h}})\big)^{T}$  and the cascaded channels ${\bm H}_k = \bm U_R {\overline{\bm H}}_k$, where ${\overline{\bm H}}_k = [{\overline{\bm H}}]_{:, (k-1)N+1:kN}$, $\forall k$.
	\end{algorithmic}
\end{algorithm}

\section{A  Two-Stage  One-bit Channel Estimation} \label{sec:two_stage}
\begin{figure}[!h]
	\centerline{\resizebox{.45\textwidth}{!}{\includegraphics{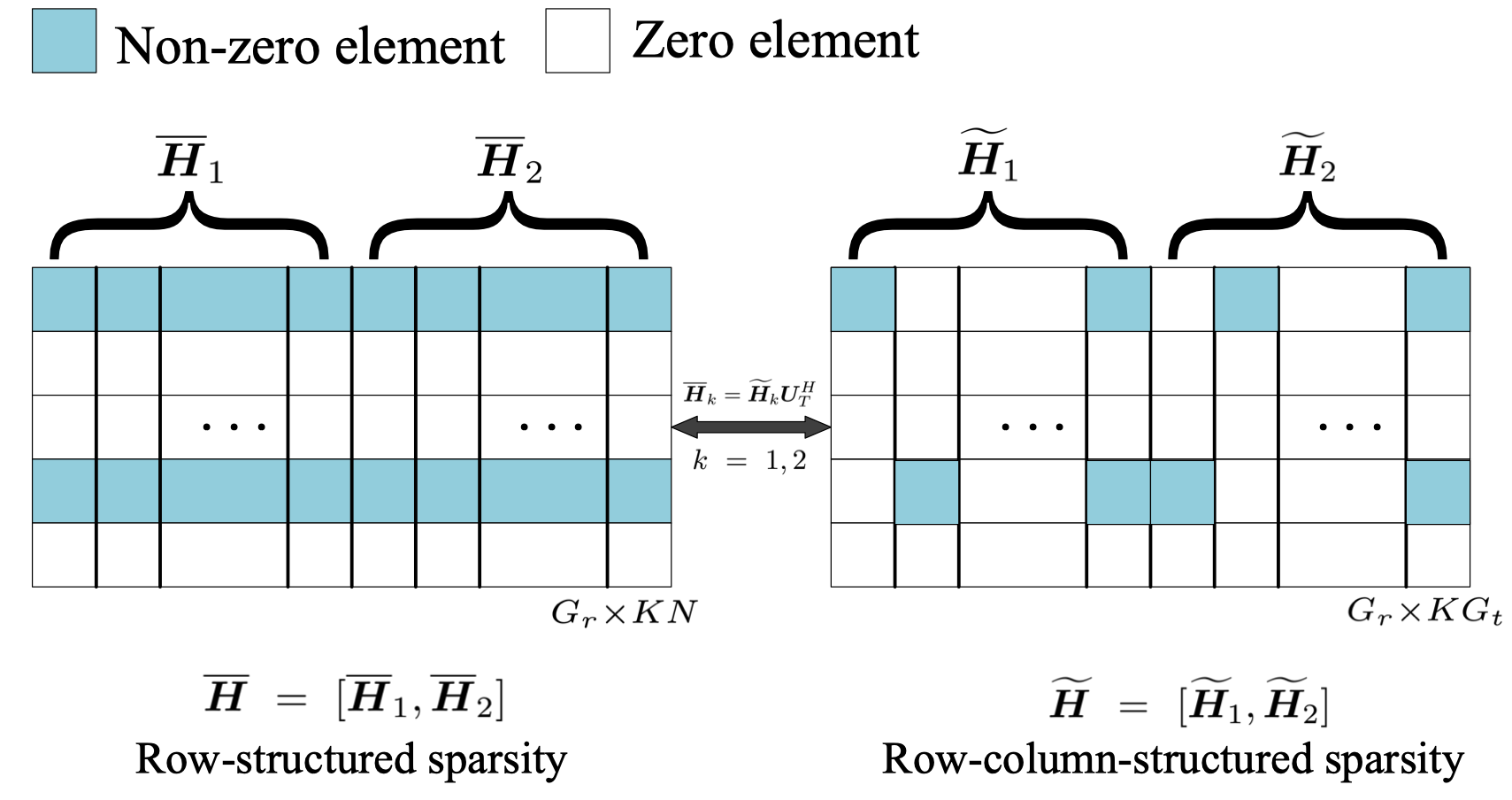}}
	}   \caption{Illustration of the sparse structure of ${\overline{\bm H}}$ and $\widetilde{\bm H}$ for $K=2$.} \label{fig:two_stage_illustration}
\end{figure}

In the last section, we have exploited the common-row sparsity of $\{ \overline{\bm H}_k\}_{k\in \cal K}$  to estimate $\{ \overline{\bm H}_k\}_{k\in \cal K}$ via the BSBL. However, if one looks into the relation between $\overline{\bm H}_k$ and $ \widetilde{\bm H}_k$, which is recapitulated below
$$\overline{\bm H}_k = \widetilde{\bm H}_k \bm U_T^H,$$
it is clear that $\widetilde{\bm H}_k$ has the same row sparsity as $\overline{\bm H}_k$, and furthermore with the limited scattering around the IRS,  $\widetilde{\bm H}_k$ should also be sparse within those non-zero rows; see Fig.~\ref{fig:two_stage_illustration} for an illustration of the two  users' case. Unfortunately, this sparsity cannot be exploited in BSBL, because $\widetilde{\bm H}_k$ multiplied with $\bm U_T^H$ has  broken  the sparse structure within the non-zero rows; see Fig.~\ref{fig:two_stage_illustration}. To exploit both the common-row sparsity of $ \{\overline{\bm H}_k \}$ and the column sparsity of $\{\widetilde{\bm H}_k\}$  within the non-zero rows, we propose a two-stage estimator in this section.

Our idea is quite simple. In the first stage, we use the BSBL estimator in Section~\ref{sec:BSBL} to estimate the common-row support $\bm \Omega$ of $\{ \overline{\bm H}_k\}$, which is also the row support of $\{\widetilde{\bm H}_k\}$. In the second stage, we turn back to the dimension-reduced SBL model (cf.~\eqref{eq: SMV_model_one_bit}):
\begin{equation} \label{eq:r_secondstage}
\bm r=\text{sgn}(\bm y)=\text{sgn}(\bm \Xi_{{\bm \Omega}} \bm h_{{\bm \Omega}} +\bm w ),
\end{equation}
where $ \bm \Xi_{\bm \Omega}  = \bm \Phi^T \otimes [\bm U_R]_{:,\bm \Omega} \in \mathbb{C}^{QM\times K |\bm \Omega| G_t}$ and $\bm h_{{\bm \Omega}} = \text{Vec}([\widetilde{\bm H}]_{\bm \Omega,:})\in \mathbb{C}^{K |\bm \Omega| G_t\times 1}$. Since the reduced channel vector $\bm h_{\bm \Omega}$ is still sparse, we can further use  the  SBL estimator in Section~\ref{sec:SBL} to estimate it. Algorithm~\ref{alg:4} summarizes the whole procedure of the two-stage  scheme.
\begin{algorithm}
	\caption{The Two-Stage One-bit Channel Estimation} \label{alg:4}
	\begin{algorithmic}[1]
		\State {\bf Input}: $\bm U_R$, $\bm \Upsilon$, $\bm \Phi$, $\bar{\bm r}$, $\bm r$, $\sigma^2$, $\gamma_{\text{th}}$ and $p_{\text{max}}$.
		\State {\bf First Stage}: Run the BSBL Algorithm~\ref{alg:3} to obtain  the hyperparameter  ${\bm \gamma}$. Then determine the row support set $\bm \Omega$ according to
		\begin{equation} \label{eq:row-support-check}
\bm \Omega  = \{n~| ~|{ \gamma}_{n}|>\gamma_{\text{th}},~n=1,\cdots, G_r\}.
		\end{equation}	
		\State {\bf Second Stage}: Choose an initial guess $\hat{\bm \alpha}_{\bm \Omega}^{(0)}=0.001\times \bm 1$, compute $\bm \Xi_{\bm \Omega}  = \bm \Phi^T \otimes [\bm U_R]_{:,\bm \Omega}$ and perform the variational EM procedure:
		\Repeat
		\State $\ell \leftarrow \ell+1$
		\State {\bf Variational E-step}: given $\hat{\bm \alpha}_{\bm \Omega}^{(\ell)}$ compute
		\State $\bm \Sigma^{(\ell)}_{h_{\bm \Omega}}  = \left({\text{Diag}(\hat{\bm \alpha}^{(\ell)}_{\bm \Omega})}^{-1} +\sigma^{-2} \bm \Xi_{\bm \Omega}^H \bm \Xi_{\bm \Omega} \right)^{-1}$
		\State Set $p=0$ and $\bm \mu^{(\ell-1)(0)}_{h_{\bm \Omega}}= \bm \mu^{(\ell-1)}_{h_{\bm \Omega}}$,	
		 \Repeat	
		 \State $p \leftarrow p+1$
		 \begin{align*}
		 \hspace{30pt}	\mu_{y,i}^{(p)} & =\frac{\sigma}{\sqrt{2}}\left(\mathfrak{R}\{r_i\} \frac{\psi (\chi_{\bm \Omega,i}^R)}{\Psi(  \chi_{\bm \Omega,i}^R )} +j\cdot \mathfrak{I}\{r_i\}\frac{\psi ( \chi_{\bm \Omega,i}^I)}{\Psi(  \chi_{\bm \Omega,i}^I )} \right) \\
		 	& +\bm \xi_{\bm \Omega, i}^T \bm \mu_{h_{\bm \Omega}}^{(\ell-1)(p-1)},~ \forall i
		 \end{align*}		 	
		 \State $\bm \mu_{h_{\bm \Omega}}^{(\ell-1)(p)}   =\sigma^{-2} \bm \Sigma^{(\ell)}_{h_{\bm \Omega}} \bm \Xi_{\bm \Omega}^H \bm \mu_{y}^{(p)}$
		 \Until{$p>p_{\text{max}}$}
		 \State $\bm \mu_{h_{\bm \Omega}}^{(\ell)}=\bm \mu_{h_{\bm \Omega}}^{(\ell-1)(p)}$.
		 \State {\bf Variational M-step}: update the hyperparemeters
		 \State $\hat{\alpha}^{(\ell+1)}_{\bm \Omega,n}=|[\bm \mu^{(\ell)}_{h_{\bm \Omega}}]_n|^2 + [\bm \Sigma^{(\ell)}_{h_{\bm \Omega}}]_{n,n},~ n=1,\ldots,K |\bm \Omega| G_t.$
		\Until{some stopping criterion is satisfied}			
	\State ${\bm h}_{\bm \Omega}= \bm \mu^{(\ell)}_{h_{\bm \Omega}}$
	\State Initialize ${\widetilde{\bm H}}=\bm 0$, $[{\widetilde{\bm H}}]_{\bm \Omega,:} = \text{Vec}^{-1}({\bm h}_{\bm \Omega})$
	\State {\bf Output}: ${\bm H}_k=\bm U_R {\widetilde{\bm H}}_k \bm U_T^H$ with ${\widetilde{\bm H}}_k = [{\widetilde{\bm H}}]_{:, (k-1)G_t+1:k G_t}$, $\forall k$.
	\end{algorithmic}
\end{algorithm}


\section{Complexity Analysis and Fast Implementation} \label{sec:complexity_fastSBL}

In this section, we analyze the computational complexity of the considered estimators and discuss a fast implementation of the SBL estimator with specially selected IRS phase shifts.
\subsection{Complexity Analysis}
For the EM-BPDN estimator, the main computation lies in calculating the gradient when performing the FISTA to solve problem~\eqref{eq:M_step2}, and the  complexity  is $\mathcal{O}(K G_r G_t Q M)$. Hence, the total computation complexity is $\mathcal{O}(\ell_{m1} K G_r G_t Q M)$, where $\ell_{m_1}$ denotes the number of EM iterations. For the SBL and the BSBL estimators, the main computation lies in calculating the matrix inverse when computing $\bm \Sigma_h$ via~\eqref{eq:mu_Sigma_b} and $\bm \Sigma_{\bar{h}}$ via~\eqref{eq:mu_Sigma_BSBL_b}. Hence, the per iteration complexity is $\mathcal{O}(K^3 G_r^3 G_t^3)$ for the SBL and $\mathcal{O}(K^3 G_r^3 N^3)$ for the BSBL. Therefore, the overall complexity of the SBL and the BSBL estimators are $\mathcal{O}(\ell_{m_2} K^3 G_r^3 G_t^3)$ and $\mathcal{O}(\ell_{m_3} K^3 G_r^3 N^3)$, resp., where $\ell_{m_2}$ and $\ell_{m_3}$ denote the number of EM iterations. Our numerical experience suggests that for  EM-BPDN, SBL and BSBL the number of EM iterations are roughly at the same order; see Fig.~\ref{fig:convergence}. Similarly, the complexity of the two-stage estimator is $\mathcal{O}(\ell_{m_3}K^3 G_r^3 N^3+\ell_{m_4} K^3 |\bm \Omega|^3 G_t^3)$, where $\mathcal{O}(\ell_{m_4} K^3 |\bm \Omega|^3 G_t^3)$ corresponds to the complexity of the second stage.


\subsection{Fast Implementation} \label{sec:fast_imp}
With carefully designed IRS phase shifts during the channel training stage, it is possible to reduce the complexity of the matrix inversion in~\eqref{eq:mu_Sigma_b} and~\eqref{eq:mu_Sigma_BSBL_b}.  Consider the case of $G_r = M$ and $G_t=N$. Then, the VAD transformation dictionaries $\bm U_R \in \mathbb{C}^{M\times M}$ and $\bm U_T\in \mathbb{C}^{N\times N}$ become unitary matrices. Let us consider a special choice of the IRS phase-shift vector $\bm \theta_q$ at the $q$-th time slot as
	\begin{equation} \label{eq:irs_phase2}
	\bm \theta_q = [\bm U_T]_{:,b_q}, \quad q=1,\cdots,Q,
	\end{equation}
	where $b_q $ is the remainder of $q$ divided by $N$. With~\eqref{eq:irs_phase2}, the costly matrix inversion in~\eqref{eq:mu_Sigma_b} can be efficiently computed by  recursively applying the block-matrix-inversion formula~\cite{zhang2017matrix}. Consequently the complexity of the matrix inversion is reduced from $\mathcal{O}(K^3 M^3 N^3)$ to $\mathcal{O}(K^3 M N)$. Similarly, the matrix inversion of the BSBL estimator in~\eqref{eq:mu_Sigma_BSBL_b} can be completed by the block diagonal matrix inversion formula, which has the computational complexity of $\mathcal{O}(M K^3 N^3)$. The specific implementation of the low-complexity recursive matrix inversion procedure and its complexity analysis are provided in the supplementary material of this paper.

\section{Numerical Results} \label{sec:num_results}
In this section, numerical results are presented to validate the effectiveness of the proposed estimators. The  normalized mean square error (NMSE) $$\frac{1}{R_m K}\sum_{r=1}^{R_m} \sum_{k=1}^K \|\bm H_k^{(r)}- \hat{\bm H}_k^{(r)} \|_F^2/\|\bm H_k^{(r)} \|_{F}^2,$$ is adopted as the performance measure, where $\bm H_k^{(r)}$, $\hat{\bm H}_k^{(r)}$ denote the true and the estimated cascaded channels for the $k$-th user at the $r$-th Monte-Carlo run and $R_m$ is the total number of runs.
To provide benchmarks, we compare the proposed methods with  the following estimation schemes:
\begin{enumerate}
\item The near maximum likelihood (nML) estimator~\cite{li2021passive}, which performs a near ML algorithm to estimate the channel with one-bit quantized measurements;
\item The Bussgang-based linear minimum mean-squared error (BLMMSE) channel estimator~\cite{Wang2021channel}, which exploits the Bussgang decomposition~\cite{li2017channel} to form a linear MMSE channel estimator for IRS-aided massive MISO systems with one-bit ADCs;
\item The support vector machine (SVM)-based channel estimator~\cite{nguyen2021svm}, which formulates the one-bit channel estimation problem as a classical SVM problem;
\item The EM-BPDN algorithm in~\cite{stockle2016channel}; see Section~\ref{sec:EM-BPDN} for the description;
\item The oracle MAP estimator (Oracle scheme), which assumes the support of the angular sparse channels $\{\widetilde{\bm H}_k\}_{k\in \cal K}$, i.e. AoA/AoDs in~\eqref{eq:cascaded_ch}, is known, and uses MAP to estimate the path gains. Clearly, the oracle MAP estimator serves as the performance lower bound for the proposed methods.
\end{enumerate}
It should be mentioned that BLMMSE and SVM  estimators both require a prior knowledge of channel correlation, however, under the IRS-assisted angular channel model, it is generally difficult to obtain the exact channel correlation. For simulation purpose, we consider two types of covariance matrices for BLMMSE and SVM: the first one follows~\cite{Wang2021channel, nguyen2021svm} and sets the covariance matrix as an identity matrix, named ``BLMMSE/SVM (identity cov.)''; the second one assumes that all the AoAs and AoDs are perfectly known (by some genie-aided schemes) and computes the covariance matrix with respect to complex gains $\zeta_{\ell_1}^G$ and $\zeta_{\ell_2,k}^r$. The resulting BLMMSE and SVM estimators are named  ``BLMMSE/SVM (genie-aided)''.

The simulation settings are as follows: The number of paths between the IRS and the BS is $L_G=2$, and the number of paths from the $k$-th user to the IRS is set as $L_{r,1}=\ldots=L_{r,K}=L_r = 6$. We assume that all
the path gains $\zeta_{\ell_1}^G$ and $\zeta_{\ell_2,k}^r$ follow ${\cal CN}(0,1)$, i.e., all users undergo similar pass loss fading~\cite{wei2021channel,chen2019channel}. The IRS reflecting elements, number of BS antennas and users are set as $N=16$ ($N_x=4$, $N_y=4$), $M=32$ and $K=3$, resp. The angular resolutions are set to $G_t=32$ ($G_{t_x}=4$, $G_{t_y}=8$)  and $G_r=64$, unless otherwise specified. We consider two simulation scenarios, i.e., with and without grid mismatch\footnote{Grid mismatch means that the true AoA/AoD does not exactly lie on the AoA/AoD grid specified by the VAD dictionaries due to limited resolution of the latter.}. For the case without grid mismatch, the AoA (resp. cascaded AoD) parameters are randomly selected from $G_r$ (resp. $G_t$) quantized angular grid points. While for the case with grid mismatch, all spatial angles are uniformly generated from $[-\pi/2,\pi/2]$. Unless otherwise specified, simulation results were obtained without grid mismatch. The pilot symbols are selected uniformly at random from the QPSK constellation with zero mean and variance $\sigma_s^2=1$ and the signal-to-noise ratio (SNR) is defined as $\sigma_s^2/\sigma^2$. Each element of the IRS reflecting matrix $\bm \Theta$ is randomly selected from the unit circle for the case of $G_r>M$ and $G_t>N$ and is set based on~\eqref{eq:irs_phase2} for the case of $G_r=M$ and $G_t=N$. The trade-off parameter $\eta$ for the EM-BPDN algorithm is fine-tuned as $\eta=0.6$ to attain overall good performance for all the tests. The stopping criterion for all the iterative schemes is either the relative successive change of the (hyper)parameters less than  $10^{-3}$ or  the maximum number of iterations $150$ reached. All the results were obtained by averaging over $500$ Monte-Carlo runs.

%

\begin{figure}[!h]
 	\vspace{-10pt}
	\centerline{\resizebox{.45\textwidth}{!}{\includegraphics{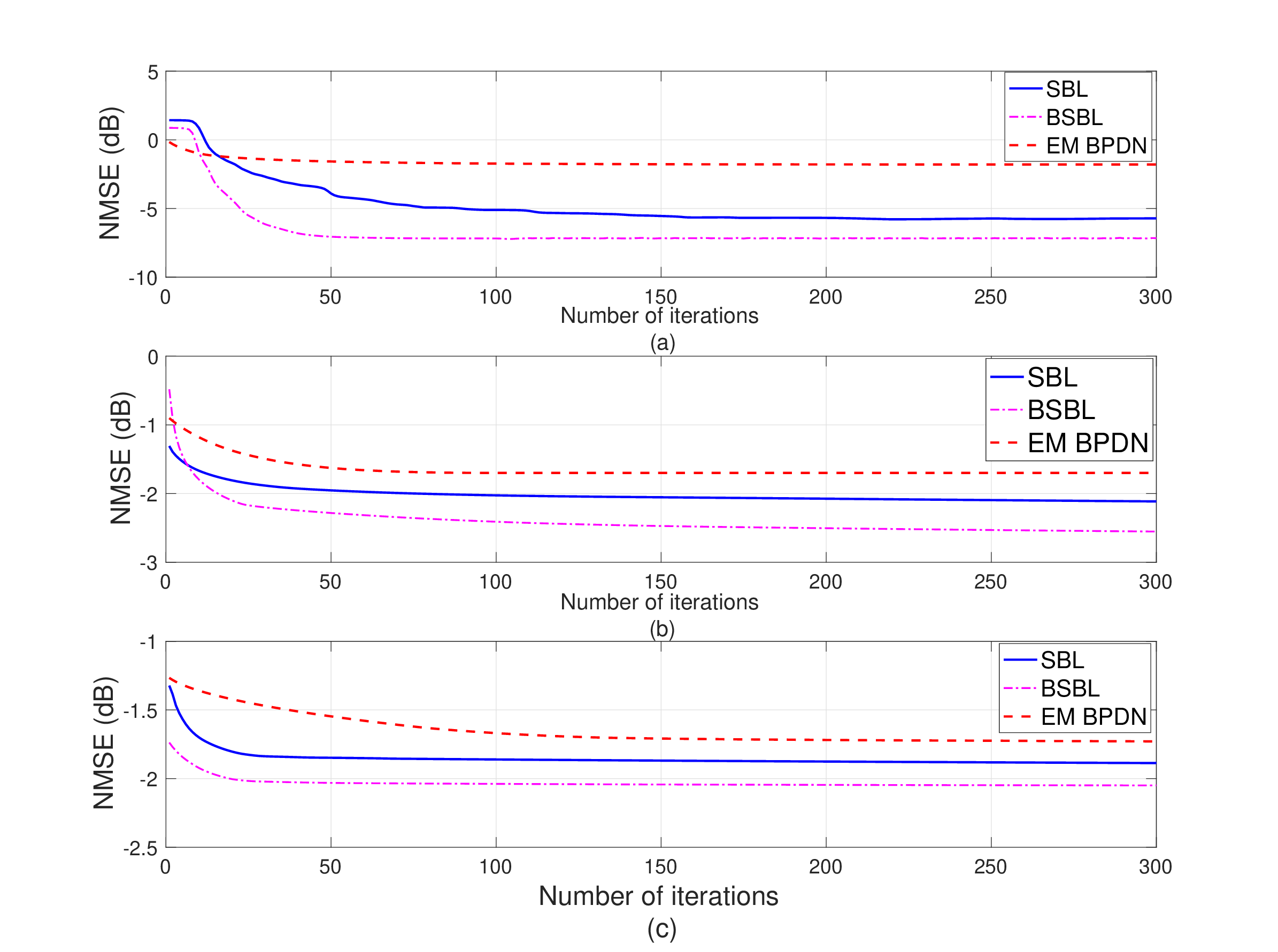}}
	}   \caption{Convergence of SBL and BSBL; (a) $\text{SNR}=0$\,dB, (b) $\text{SNR}=15$\,dB, (c) $\text{SNR}=30$\,dB. ($Q=88$)} \label{fig:convergence}
\end{figure}

First, let us check the convergence behaviors of the proposed schemes. Fig.~\ref{fig:convergence} shows the convergence results of the EM-BPDN, the SBL and the BSBL schemes for $\text{SNR}=0$\,dB, $\text{SNR}=15$\,dB and $\text{SNR}=30$\,dB. It is seen that the NMSEs of all the schemes tend to be flattened after 150 iterations for different SNRs, and that the convergence speed of the SBL and the BSBL is generally comparable---the BSBL is slightly faster than the SBL for low SNR, and the situation is reversed when SNR increases to 15\,dB, but for high SNR, both algorithms have nearly the same convergence speed.

 \begin{figure}[!h]
	 	\vspace{-10pt}
	\centerline{\resizebox{.45\textwidth}{!}{\includegraphics{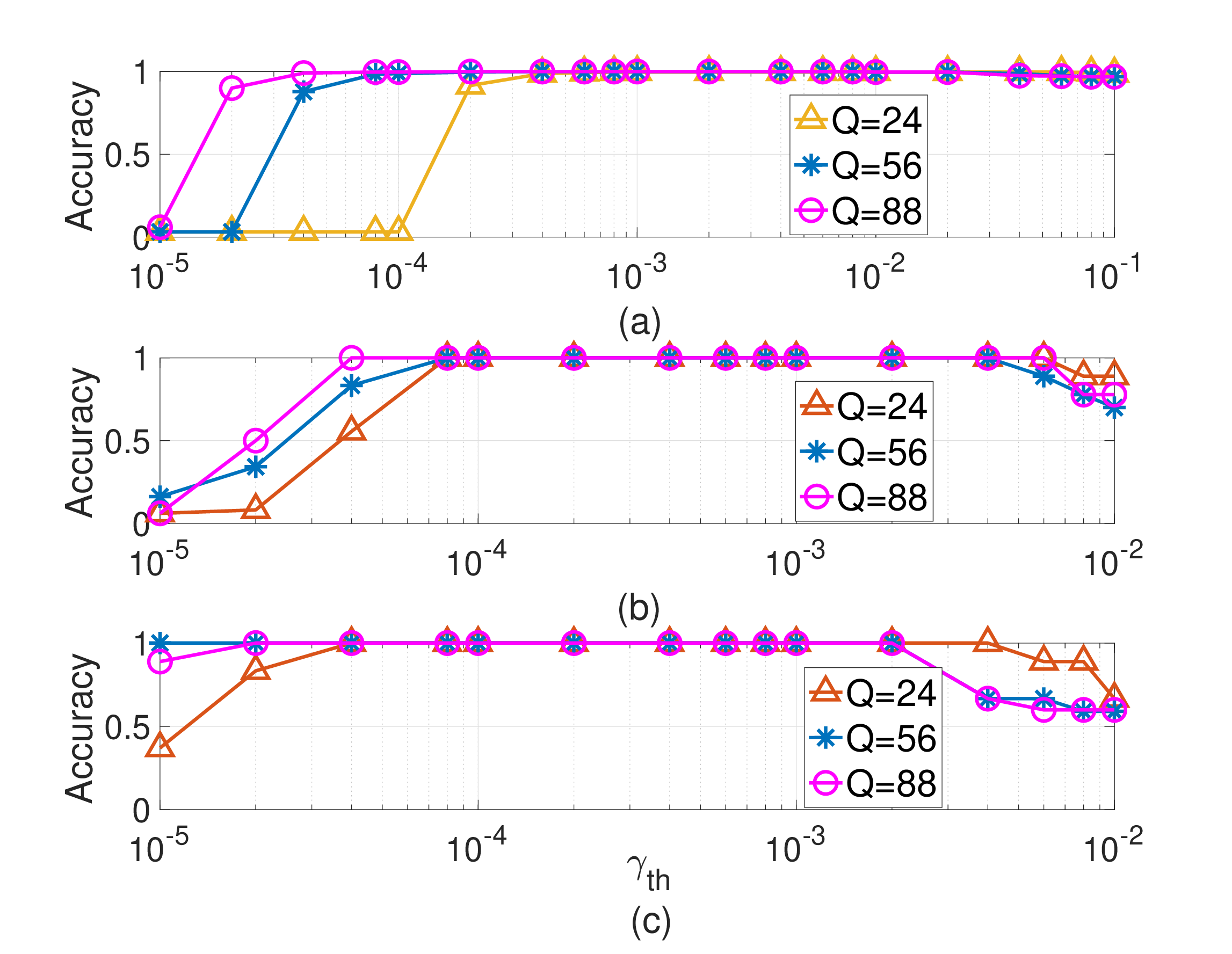}}
	}   \caption{Impact of the threshold $\gamma_{\text{th}}$ on the row-support recovery accuracy for different training overheads $Q$; (a) $\text{SNR}=0$\,dB, (b) $\text{SNR}=15$\,dB, (c) $\text{SNR}=30$\,dB.} \label{fig:accuracy_vs_gammath}
\end{figure}

Next, let us investigate the impact of  the threshold $\gamma_{\text{th}}$ [cf.~\eqref{eq:row-support-check}] on the  row-support recovery in the two-stage scheme. Herein, we adopt the widely used ``accuracy'' as the recovery performance metric, which
takes both the false alarm and the miss detection into account and is formally defined as $\text{Accuracy}=\frac{\text{TP}+\text{TN}}{\text{TP}+\text{FA}+\text{TN}+\text{MD}}$ (TP: true positive, TN: true negative, FA: false alarm, MD: miss detection)~\cite{fawcett2006introduction}.	Fig.~\ref{fig:accuracy_vs_gammath} shows row-support recovery accuracy versus the threshold $\gamma_{\text{th}}$ for different length of training overhead $Q$ when $\text{SNR}=0$\,dB, $\text{SNR}=15$\,dB and $\text{SNR}=30$\,dB. We can observe from the figures that perfect recovery can be attained within a wide range of $\gamma_{\text{th}}$, e.g., $10^{-4}\leq  \gamma_{\text{th}} \leq 10^{-2}$. This suggests that the proposed two-stage scheme is less sensitive to the choice of the threshold $\gamma_{\text{th}}$ for both low and high SNRs. In light of the results in Fig.~\ref{fig:accuracy_vs_gammath}, we set $\gamma_{\text{th}} = 10^{-3}$ in the following simulations.

\begin{figure}[!h]
 	\vspace{-10pt}
	\centerline{\resizebox{.43\textwidth}{!}{\includegraphics{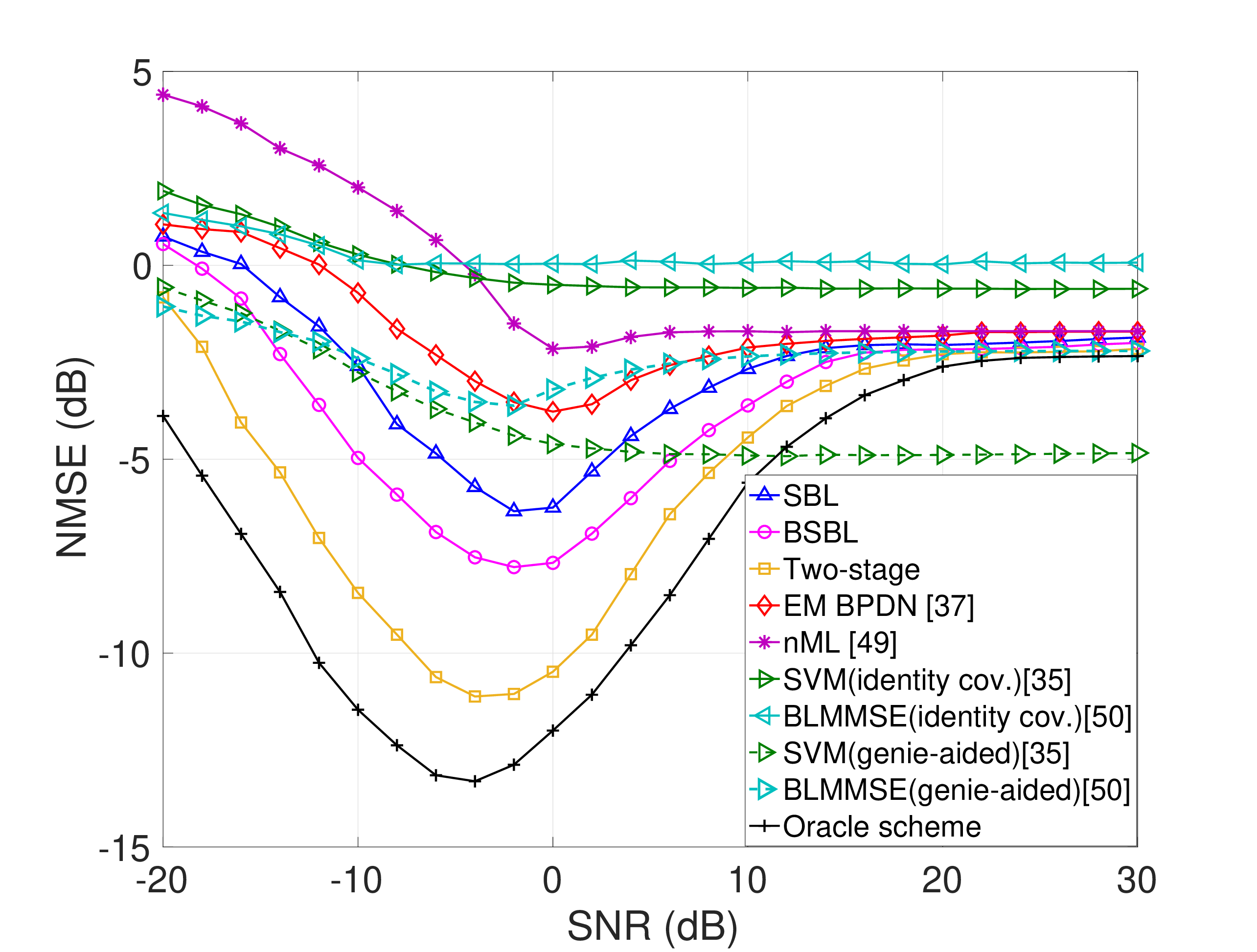}}
	}   \caption{NMSE vs. SNR ($Q=88$).} \label{fig:MSE_vs_SNR}
\end{figure}

Fig.~\ref{fig:MSE_vs_SNR} shows the NMSE performance versus the SNR for $Q=88$. From the figure, we have the following observations. Firstly, the proposed three estimators outperform the benchmark schemes  for all the SNRs tested, and better NMSE is attained as more structured sparsity is utilized. We should point out that the  performance of EM-BPDN  relies on the trade-off parameter $\eta$, which  needs to be carefully tuned in advance; therefore, it is generally hard to attain overall good performance with a fixed $\eta$. 
Secondly, the BLMMSE (identity cov.) and SVM (identity cov.) suffer severe performance degradation and even perform worse than the nML at high SNRs. The reason for this phenomenon is that the BLMMSE and SVM  both require prior knowledge of the channel covariance matrix, while for our considered IRS-aided cascade angular channel (cf.~Eqn.~\eqref{eq:cascaded_ch}), it is difficult, if not impossible, to specify a reasonably good channel covariance matrix in practice (identity covariance matrix may not be the best choice for the IRS case). Therefore, the mismatch between these estimators' prerequisites and the IRS channel characteristics leads to poor performance for the BLMMSE and SVM estimators. Moreover, with genie-aided covariance matrix, the performances of BLMMSE (genie-aided) and SVM (genie-aided) are improved, but they are still inferior to our proposed methods at low and middle SNRs, because BLMMSE and SVM are not able to fully exploit the channel sparsity. Finally, the NMSEs of the proposed estimators first decrease and then increase as SNR grows. This is a typical phenomenon termed stochastic resonance~\cite{mo2014channel} for systems with quantization. By contrast, the SVM (genie-aided) scheme is more robust against the stochastic resonance  and outperforms the proposed methods at high SNRs; this may be attributed to the use of exact channel correlation information, which may alleviate the stochastic resonance  to some extent.

\begin{figure}
	\vspace{-10pt}
		\centerline{\resizebox{.43\textwidth}{!}{\includegraphics{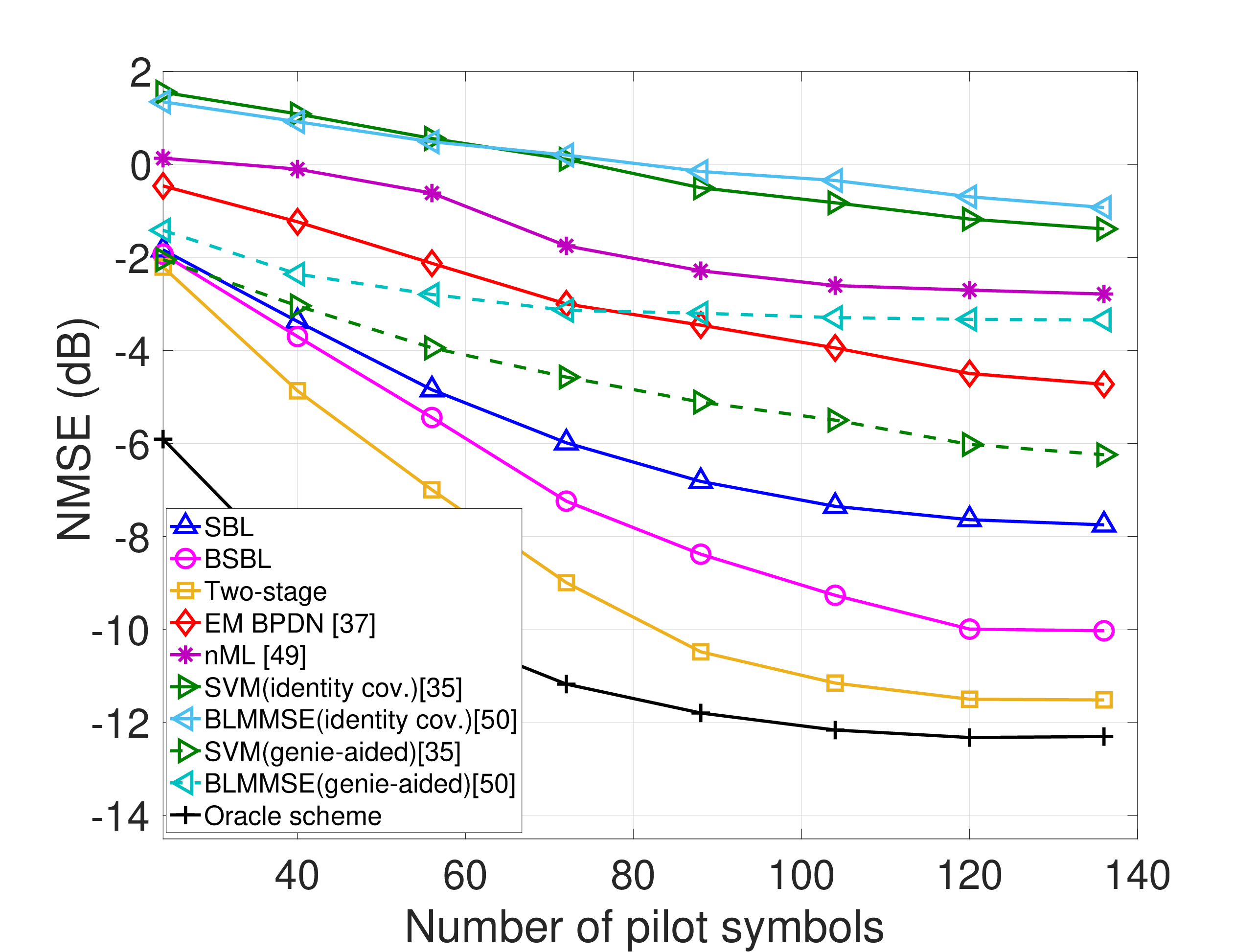}}}	
 	\caption{NMSE  vs. training overhead $Q$.}\label{fig:MSE_vs_Qa}
\end{figure}

Fig.~\ref{fig:MSE_vs_Qa} shows the impact of the training overhead $Q$ on the NMSE at SNR$=0$\,dB. From the two figures, we see that the nML, BLMMSE and SVM schemes, which do not exploit the channel sparsity, perform poorly in comparison to the proposed schemes. In particular, our proposed schemes require only about 24 pilot symbols to achieve an NMSE of $-2$\,dB, while  the nML scheme requires about 72 pilot symbols, and the BLMMSE (identity cov.) and SVM (identity cov.) schemes require more than 140 pilot symbols to achieve an NMSE of $-2$\,dB. This result suggests that the training overhead may be reduced by leveraging the sparsity of the  cascaded channels. We also observe that the two-stage scheme attains the best NMSE among the three proposed SBL schemes. Specifically, to achieve an NMSE of $-7$\,dB, the required number of pilot symbols is approximately 104, 72 and 56 for the proposed SBL, BSBL and two-stage schemes, resp. This again
verifies that the more diverse sparsity is exploited, the less training overhead is needed.



%

\begin{figure}[!h]
 		\vspace{-10pt}
	\centerline{\resizebox{.43\textwidth}{!}{\includegraphics{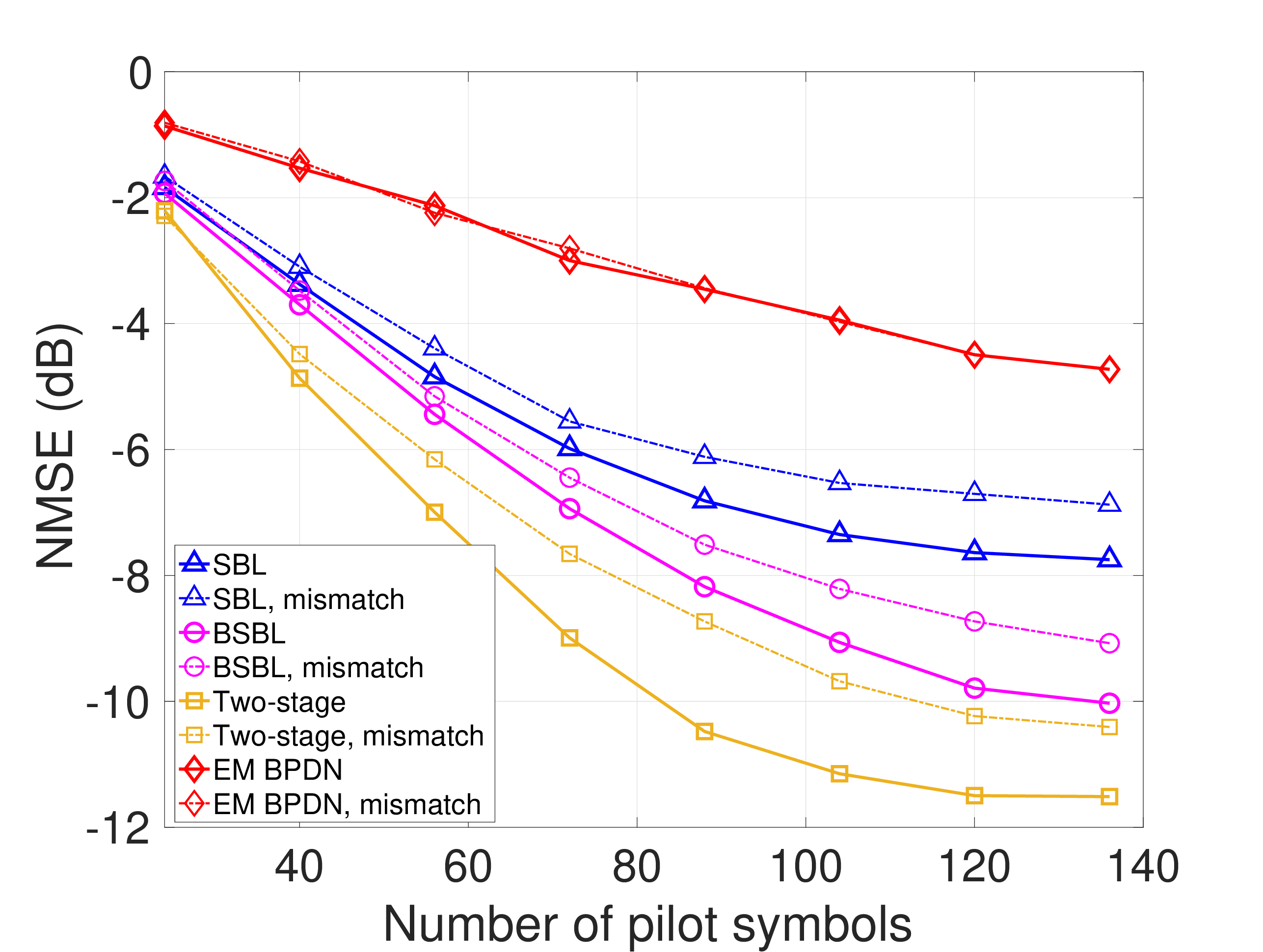}}
	}   \caption{NMSE  vs.  $Q$ with grid mismatch.} \label{fig:mismatch_SNR0}
\end{figure}

In Fig.~\ref{fig:mismatch_SNR0}, we investigate the sensitivity of the proposed sparse channel estimation schemes to the grid mismatch at SNR$=0$\,dB. We plot the NMSE of different algorithms as a function of $Q$ by considering two scenarios, i.e., with and without gird mismatch. It can be observed that when there is grid mismatch,  EM-BPDN  has negligible  performance degradation, whereas the SBL-based schemes are more sensitive  to the grid mismatch, especially the two-stage scheme. The reason for this is that in the presence of grid mismatch, the number of nonzero elements in $\widetilde{\bm H}_k$ will expand due to  power leakage; consequently,
it is hard to correctly detect the row support and the estimation in the second stage  may be misguided by the wrong  support. Nevertheless, the two-stage estimation scheme still shows a large improvement in NMSE as compared with the  other schemes even in the presence of grid mismatch.


\begin{figure}[!h]
 		\vspace{-10pt}
	\centerline{\resizebox{.43\textwidth}{!}{\includegraphics{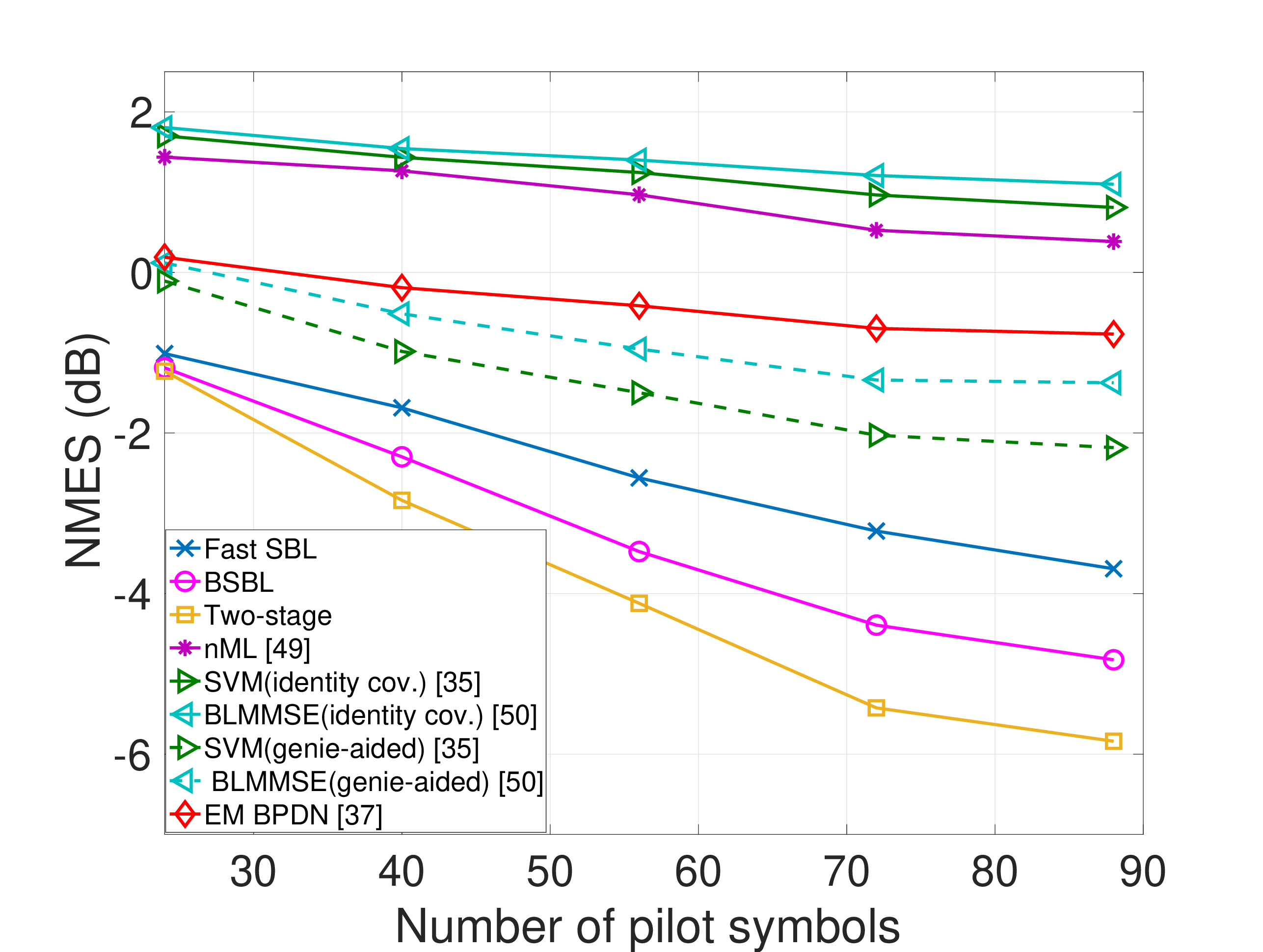}}
	}   \caption{NMSE  vs.  $Q$ for $M=N=32$.} \label{fig:fastSBL_SNR0}
\end{figure}

Next, we show the effectiveness of the fast implementation of the SBL estimator. The setting is basically the same as Fig.~\ref{fig:mismatch_SNR0} except that $G_r=M$ and $G_t = N$ are set  to make the VAD transformation dictionaries unitary; cf.~Sec.~\ref{sec:fast_imp}. We should mention that under such setting, the resolution of the VAD transformation dictionaries is limited by the  size of  arrays at the BS and the IRS.  Also, grid mismatch is considered.  Fig.~\ref{fig:fastSBL_SNR0}  shows the NMSE performance against the pilot overhead $Q$. We can see that even though the VAD transformation dictionaries are not super-resolution and  the grid mismatch exists, the proposed SBL-based estimators still outperform other benchmark schemes. 

\begin{figure}[!h]
 	\vspace{-10pt}
	\centerline{\resizebox{.43\textwidth}{!}{\includegraphics{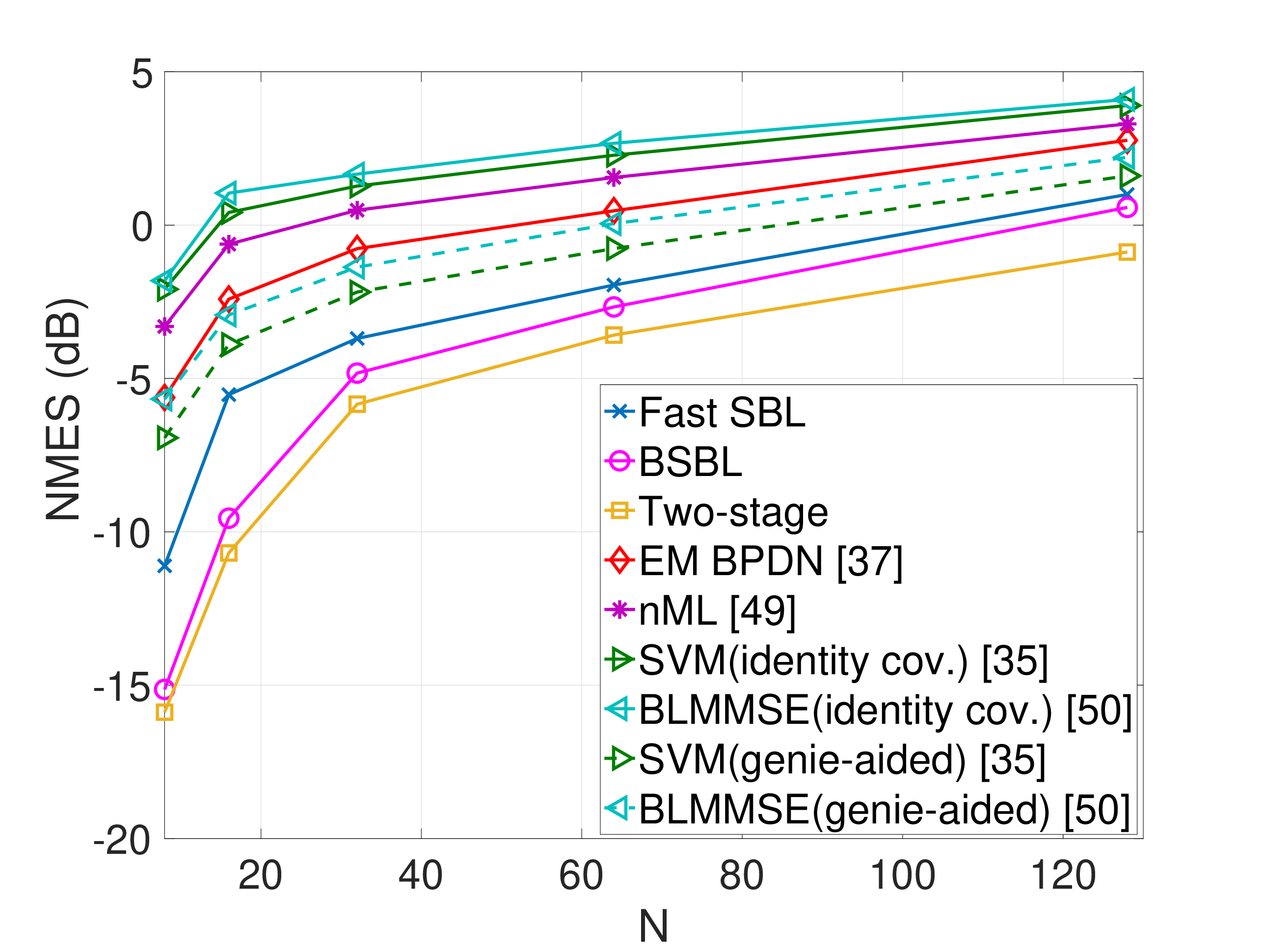}}
	}   \caption{NMSE vs. $N$ for $M=32$ and $Q=88$.} \label{fig:MSE_vs_N}
\end{figure}

Fig.~\ref{fig:MSE_vs_N} shows the impact of the number of IRS elements $N$ on the NMSE performance for $M=32$ and $Q = 88$. Other simulation settings are the same as Fig.~\ref{fig:fastSBL_SNR0}. As seen, the NMSE of all estimation schemes increases as the number of IRS elements grows. And the proposed schemes outperform the other schemes over the whole range of $N$ under the considered setting. 
  

\begin{figure}[!h]
 		\vspace{-10pt}
	\centerline{\resizebox{.43\textwidth}{!}{\includegraphics{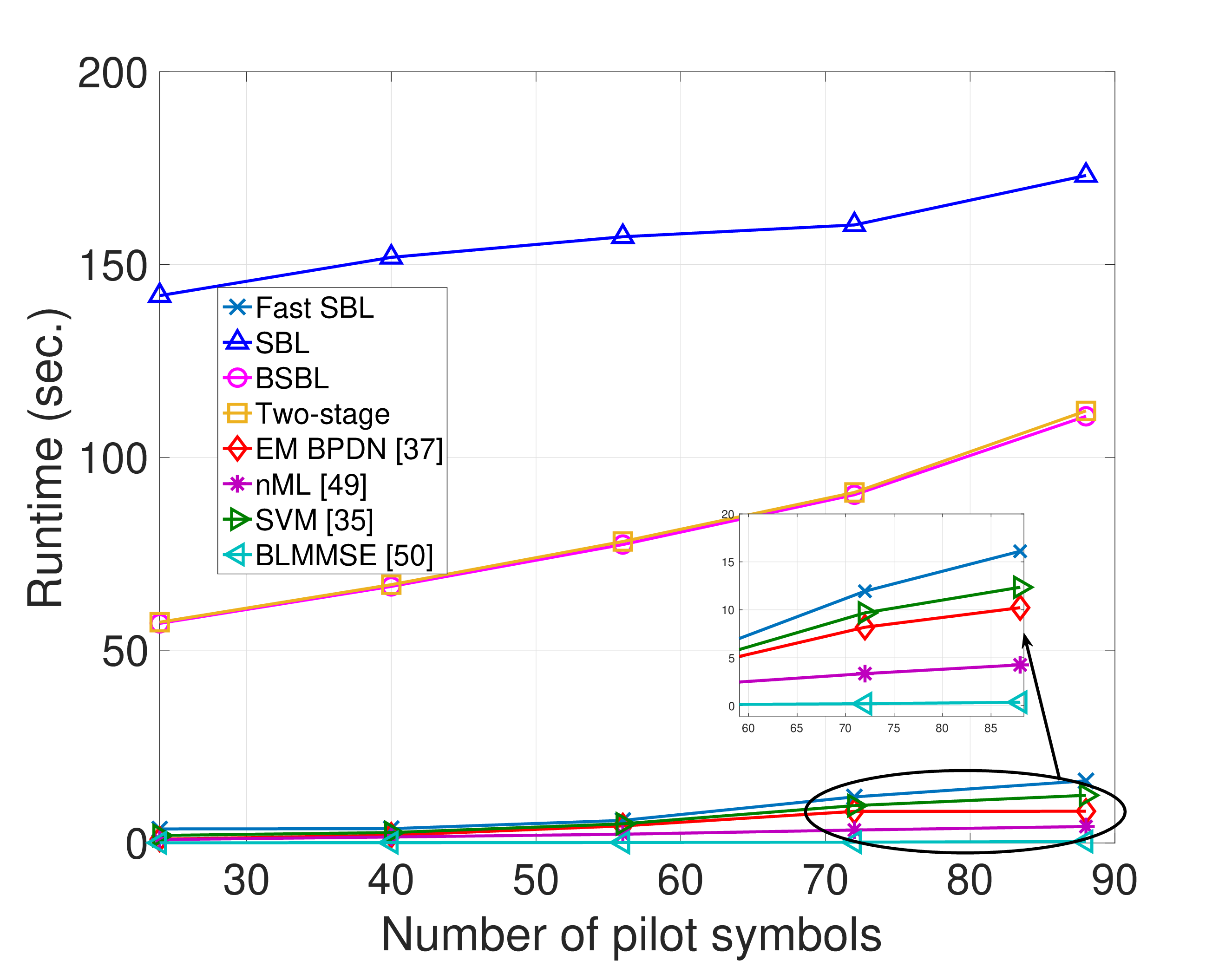}}
	}   \caption{The runtime vs. $Q$ for $M=N=32$ and $\text{SNR}=0$\,dB.} \label{fig:runtime}
	\vspace*{-1.4\baselineskip}
\end{figure}

Fig.~\ref{fig:runtime} plots the runtime  of different schemes under the same setting as Fig.~\ref{fig:fastSBL_SNR0}. We see that the runtime of the ``Fast SBL'' scheme is far less than the original SBL scheme, and is comparable to EM-BPDN, especially when the number of pilots is small. Moreover, the runtime of the ``Two-stage'' scheme is almost the same as that of the BSBL scheme, since the fast implementation of the SBL is applied in the second stage of the ``Two-stage'' scheme.

\section{Conclusions} \label{sec:conclusions}
In this paper, we have studied the  uplink channel estimation for the IRS-aided MU-MISO mmWave system with one-bit quantization at the BS. By exploiting the structured sparsity of the IRS channel, we have developed three estimators under the SBL framework. The proposed three  estimators utilize different levels of the structured sparsity, including the elementwise sparsity, the common row sparsity and the row-column sparsity. In general, the more diverse structured sparsity is exploited, the better estimation performance can be achieved. There are some directions worth further investigation. First, the numerical results revealed that the SBL-based methods are  sensitive to the grid mismatch.  As a future work, it is worth taking the grid mismatch into the design of the SBL-based estimators. For example, one may adaptively learn the VAD transformation dictionaries $\bm U_R$ and $\bm U_T$ along with the update of the hyperparameters, or employ multi-level resolution dictionaries with non-uniform angular grids to alleviate the effect of grid mismatch.  Secondly, the joint estimation of channel parameters and noise variance. In this work, we have assumed that noise variance is known, but in practice the noise variance needs to be estimated. It is worth investigating how to extend the proposed methods for unknown noise variance case. Finally, the joint design of the pilot symbols, IRS phase-shift and the channel estimator. In Sec.~\ref{sec:fast_imp}, we gave a fast implementation of the SBL for the special case $G_r = M$ and $G_t = N$. It is worth further investigating how to jointly  design  the pilot symbols and the IRS phase-shift to further reduce the complexity of the SBL estimators under a more general scenario setting. In addition, in our previous work~\cite{wang2020one}, we have considered the joint design of the  one-bit transmit symbols and the IRS phase-shift. It would be also interesting to  incorporate the design method in~\cite{wang2020one} into the current work to further improve the estimation performance.

 \newpage
 \title{Supplementary Material of ``One-Bit Channel Estimation for IRS-aided
 	Millimeter-Wave Massive MU-MISO System''}

 \author{ Silei Wang,  Qiang Li and Jingran Lin
 }
 \maketitle
	\section{Calculating $\mathbb{E}_{q(\bm y)}[\bm y]$}
This section gives the detailed derivation of calculating $\mathbb{E}_{q(\bm y)}[\bm y]$. Firstly, we observe from~(29) that the right-hand side of the last equation can be rewritten as
\begin{equation}
\ln q(\bm y)=\sum_{i=1}^{Q M}\ln q(y_{i}) \tag{54}
\end{equation}
where
\begin{equation}\label{eq:q_y_i}
\ln  q(y_{i})\hspace{-3pt} =\hspace{-3pt}\ln p(r_{i}|y_{i})-\frac{|y_{i} - \bm \xi_{i}^T \mathbb{E}_{q(\bm h)}[\bm h] |^2}{\sigma^2}+ \text{const}. \tag{55}	
\end{equation}
Therefore, we can evaluate $\mathbb{E}_{q(\bm y)}[\bm y]$ by separately evaluating $\mathbb{E}_{q(y_{i})}[y_i]$ for $i=1,\cdots, QM$.

Denote $y_{R,i}=\mathfrak{R}\{y_i \}$, $y_{I,i}=\mathfrak{I}\{y_i \}$. We  express $\mathbb{E}_{q(y_{i})}[y_i]$ as
\begin{equation} \label{eq: Eqy_y}
\mathbb{E}_{q(y_{i})}[y_i] = \mathbb{E}_{q(y_{R,i})}[y_{R,i}] + j \mathbb{E}_{q(y_{I,i})}[ y_{I,i}]. \tag{56}
\end{equation}
Next, we will focus on calculating the real part $\mathbb{E}_{q(y_{R,i})}[y_{R,i}]$ and the imaginary part $\mathbb{E}_{q(y_{I,i})}[ y_{I,i}]$  can be calculated similarly.
Let $r_{R,i}=\mathfrak{R}\{r_i \}$, $z_i \triangleq \bm \xi_{i}^T \mathbb{E}_{q(\bm h)}[\bm h]$ and $z_{R,i}=\mathfrak{R}\{z_i \}$, we have
\begin{equation}\label{eq:q_yR_i}
\ln  q(y_{R, i})\hspace{-3pt} =\hspace{-3pt}\ln p(r_{R,i}|y_{R,i})-\frac{(y_{R,i} - z_{R,i})^2}{2 (\sigma^2/2)}+ \text{const}. \tag{57}	
\end{equation}
Take the exponential on both sides of~\eqref{eq:q_yR_i} and normalize it to get
\begin{equation*}  \label{eq:qy_n}
\begin{aligned}
q(y_{R,i}) &=\frac{ p(r_{R,i}|y_{R,i}) \exp(-\frac{(y_{R,i} - z_{R,i})^2}{2 (\sigma^2/2)})}{\int p(r_{R,i}|y_{R,i}) \exp(-\frac{(y_{R,i} - z_{R,i})^2}{2 (\sigma^2/2)}) d y_{R,i} }  \\
& =\frac{ p(r_{R,i}|y_{R,i}) \mathcal{N} \big(y_{R,i}|z_{R,i}, \sigma^2/2\big) }{\int p(r_{R,i}|y_{R,i}) \mathcal{N} \big(y_{R,i}|z_{R,i}, \sigma^2/2\big) d y_{R,i} } \\
&= \frac{p(r_{R,i}|y_{R,i})\psi \left(\frac{y_{R,i}-z_{R,i}}{\sqrt{\sigma^2/2}} \right) }{\int p(r_{R,i}|y_{R,i})\psi \left(\frac{y_{R,i}-z_{R,i}}{\sqrt{\sigma^2/2}} \right) d y_{R,i}}
\end{aligned}
\end{equation*}
for $i=1,\ldots, QM$. Therefore, the expectation $\mathbb{E}_{q(y_{R,i})}[y_{R,i}]$ is given by
\begin{equation} \label{eq:cal_Ey}
\begin{aligned}
\hspace{-5pt} \mathbb{E}_{q(y_{R,i})}[y_{R,i}]& = \int y_{R,i} q(y_{R,i}) d y_{R,i} \\
& = \hspace{-3pt}\frac{\int y_{R,i}p(r_{R,i}|y_{R,i})\psi \left(\frac{y_{R,i}-z_{R,i}}{\sqrt{\sigma^2/2}} \right) d y_{R,i}}{\int p(r_{R,i}|y_{R,i})\psi \left(\frac{y_{R,i}-z_{R,i}}{\sqrt{\sigma^2/2}} \right) d y_{R,i}}.
\end{aligned}\tag{58}
\end{equation}
With $r_{R,i}=\mathfrak{R}\{r_i \} = \text{sgn}(y_{R,i})$, the conditional probability of $r_{R,i}$ given $y_{R,i}$ can be expressed as
\begin{equation*}
p(r_{R,i}|y_{R,i})\hspace{-2pt}=\hspace{-2pt}\left\{
\begin{aligned}
1,~ & r_{R,i} = \text{sgn}(y_{R,i}) \\
0, ~& \text{otherwise}
\end{aligned}
\right.	\hspace{-5pt} =\hspace{-2pt}\left\{
\begin{aligned}
1, ~& y_{R,i} \in (r_i^{\text{low}},r_i^{\text{up}}] \\
0, ~& \text{otherwise}
\end{aligned}
\right.
\end{equation*}
with $r_i^{\text{low}}\hspace{-4pt} =\hspace{-2pt}\left\{
\begin{aligned}
& - \infty, r_{R,i} = -1 \\
& 0, ~~~~~ r_{R,i} = +1
\end{aligned}
\right.$ and $r_i^{\text{up}}\hspace{-4pt} =\hspace{-2pt}\left\{
\begin{aligned}
& 0, ~~~~~ r_{R,i} = -1 \\
& +\infty, ~ r_{R,i} = +1
\end{aligned}
\right.$.
Then, the expectation in~\eqref{eq:cal_Ey} can be evaluated as
\begin{equation} \label{eq:E_qyR_yR}
\begin{aligned}
\mathbb{E}_{q(y_{R,i})}[y_{R,i}] & = \frac{\int_{r_i^{\text{low}}}^{r_i^{\text{up}}} y_{R,i}\psi \left(\frac{y_{R,i}-z_{R,i}}{\sqrt{\sigma^2/2}} \right) d y_{R,i}}{\int_{r_i^{\text{low}}}^{r_i^{\text{up}}} \psi \left(\frac{y_{R,i}-z_{R,i}}{\sqrt{\sigma^2/2}} \right) d y_{R,i}}  \\
& = r_{R,i} \frac{\sigma}{\sqrt{2}} \frac{\psi ( \chi_i^R)}{\Psi(  \chi_i^R )} + z_{R,i}
\end{aligned}\tag{59}
\end{equation}
for $i=1,\ldots, QM$, where $\chi_i^R = \frac{r_{R,i} z_{R,i}}{\sqrt{\sigma^2/2}}$,$\psi(x)=\frac{1}{\sqrt{2\pi}} \exp(-\frac{x^2}{2})$ and $\Psi(x) = \int_{-\infty}^{x} \psi (t) dt$.

Similarly, the imaginary part $\mathbb{E}_{q(y_{I,i})}[ y_{I,i}]$ has the same form as~\eqref{eq:E_qyR_yR}, with all ``$\mathfrak{R}$'' and  ``$R$'' replaced with ``$\mathfrak{I}$'' and ``$I$'', resp. Therefore, the expectation in~\eqref{eq: Eqy_y} can be expressed as
\begin{equation}
\mathbb{E}_{q(y_{i})}[y_i] = \frac{\sigma}{\sqrt{2}}\left(r_{R,i}\frac{\psi ( \chi_i^R)}{\Psi(  \chi_i^R )} +j\cdot r_{I,i}\frac{\psi ( \chi_i^I)}{\Psi(  \chi_i^I )} \right) +z_i \tag{60}
\end{equation}
for $i=1,\ldots, QM$.

\section{ One-Bit IRS Channel Estimation With Direct Channel}
In this section, we show that the proposed SBL estimator can be easily extended to the case with direct link between users and the BS. To be specific, with the direct link,  the received signal model is modified as
\begin{equation} \label{eq: y_q1}
\bm y_q = \sum_{k=1}^{K} (\bm H_k \bm \theta_q + \bm h_{d,k} ) s_{q,k} +\bm w_q, ~ q=1,\ldots, Q, \tag{61}
\end{equation}
where $\bm h_{d,k}$ denotes the direct channel from user $k$ to the BS. By using the virtual angular-domain (VAD) representations of $\bm H_k$ and $\bm h_{d,k}$
\begin{equation} \label{eq:VAD_expression}
\bm H_k = \bm U_R \widetilde{\bm H}_k \bm U_T^H,~~\bm h_{d,k}=\bm U_R \tilde{\bm h}_{d,k}, \tag{62}
\end{equation}
we re-express $\bm y_q$ as
\begin{equation}\label{eq: y_q2}
\bm y_q  = \bm U_R \big(\sum_{k=1}^{K} \widetilde{\bm H}_k s_{q,k} \big) \bm U_T^H \bm \theta_q + \bm U_R \sum_{k=1}^{K} \tilde{\bm h}_{d,k} s_{q,k} + \bm w_q \tag{63}
\end{equation}
for $q=1,\ldots, Q$. By collecting $\bm Y = [\bm y_1, \ldots, \bm y_Q]$ and vectorizing $\bm Y$, we obtain
\begin{equation}
\begin{aligned}
\bm y & =  {\rm Vec}(\bm Y) \\
& = \bm \Xi \bm h +  \bm \Xi_{ d} \bm h_{ d} + \bm w \\
& =\tilde{ \bm \Xi } \tilde{\bm h} + \bm w 
\end{aligned}\tag{64}
\end{equation}
where $\tilde{ \bm \Xi } = [\bm \Xi, ~ \bm \Xi_d]$, $\tilde{\bm h}  =  [\bm h^T, ~\bm h_d^T]^T$, $\bm h$ and $\bm \Xi$ are defined in (13),  $\bm \Xi_d = [\bm s_1, \ldots, \bm s_Q]^T \otimes \bm U_R$ and $\bm h_d = [\bm h_{d,1}^T, \ldots, \bm h_{d, K}^T]^T$ are introduced by the direct channels. Clearly, if we consider element-wise sparsity of $\tilde{ \bm h}$,  the proposed SBL algorithm can be directly applied to estimate $\tilde{ \bm h}$ from the one-bit quantization of $\bm y$.

\section{Fast SBL Implementation and Complexity}

This section gives the specific implementation of the low-complexity recursive matrix inversion procedure for fast SBL in Section VI-B of the main manuscript, and analyze its complexity.


For the case of $G_r=M$ and $G_t=N$, the VAD transformation dictionaries $\bm U_R \in \mathbb{C}^{M\times M}$ and $\bm U_T\in \mathbb{C}^{N\times N}$ become unitary matrices. Consider the special design of the IRS phase-shift vector $\bm \theta_q$ at the $q$-th time slot as
\begin{equation} \label{eq: theta_choose}
\bm \theta_q = [\bm U_T]_{:,b_q}, q=1,\cdots,Q, \tag{65}
\end{equation}
where $b_q$ is the remainder when $q$ is divided by $N$. With~\eqref{eq: theta_choose}, the costly matrix inversion $(\text{Diag}(\bm \alpha)^{-1} + \sigma^{-2} \bm \Xi^H \bm \Xi)^{-1}$ in the E-step of SBL can be efficiently computed by recursively applying the block-matrix-inversion formula. Specifically, using~\eqref{eq: theta_choose}, one can verify that the matrix the matrix $\bm E \triangleq  \text{Diag}(\bm \alpha)^{-1} + \sigma^{-2} \bm \Xi^H \bm \Xi \in 
\mathbb{C}^{KMN\times KMN}$ takes the following special block form: 
\begin{align*}
\bm E =\begin{bmatrix}
\bm E_{11},\bm E_{12},\cdots,\bm E_{1K}\\
\bm E_{21},\bm E_{22},\cdots,\bm E_{2K}\\
\cdots \\
\bm E_{K1},\bm E_{K2},\cdots,\bm E_{KK}
\end{bmatrix},
\end{align*}
in which $\bm E_{ij}$ is a diagonal matrix given by~\eqref{eq:E_def},
\begin{figure*}[!t]
	\setlength\arraycolsep{0pt}
	\begin{equation}\label{eq:E_def}
	\bm E_{ij} = \begin{cases}
	[\text{Diag}(\bm \alpha)^{-1}]_{(i-1)MN+1:iMN} + \sigma^{-2} \text{Blkdiag} \left( (a+1) \bm I_{b}, a \bm I_{N-b}\right) \otimes \bm I_M, & {\rm for}~ i=j \\
	\sigma^{-2}\text{Diag}\left(\sum_{m=0}^{a} [\bar{\bm s}_{i,j}]_{mN+1:(m+1)N}\right)\otimes \bm I_M, & {\rm for}~ i \neq j
	\end{cases} \tag{66}
	\end{equation} 
	
	\hrulefill
\end{figure*}
where $\bar{\bm s}_{i,j} = [{\bm s}_{i,j};\bm 0]\in \mathbb{C}^{(a+1)N\times 1}$ and ${\bm s}_{i,j} = \bm s_i^* \odot \bm s_j\in \mathbb{C}^{Q\times 1}$; $a$ and $b$ denote the modulus and the remainder when $Q$ is divided by $N$, i.e., $Q=aN + b$; $\bm s_i\in \mathbb{C}^{Q\times 1}$ denotes the pilot sequence sent by the $i$-th user during $Q$ time slots and $\bm s_i^*$ is the conjugate of the vector $\bm s_i$; $\text{Blkdiag} (\bm A, \bm B)$ denotes the block diagonal matrix with the diagonal blocks being $\bm A$ and $\bm B$; $\bm s_i \odot \bm s_j$ denotes the dot product of two vectors $\bm s_i$ and $\bm s_j$. Obviously, each $\bm E_{ij}\in \mathbb{C}^{MN\times MN}$, $\forall i,j$ is an invertible diagonal matrix and $\bm E_{ij} =\bm E_{ji}^H$. Therefore, we can compute $\bm E^{-1}$ by recursively applying the block-matrix inversion formula~\cite{ZhangXD}. Algorithm~\ref{alg:recursive}, which follows a divide-and-conquer idea, summarizes the recursive inverse procedure. Lines 1-3 test for the base case, where the sub-matrix is an $MN \times MN$ diagonal matrix. Note that $\text{Diag}(\bm E)$ denotes a vector with elements being the diagonal elements of the matrix $\bm E$ and $\text{Diag}(\bm e)$ represents a diagonal matrix with diagonal elements being $\bm e$. The symbol ``./'' denotes the element-wise division, i.e., 1./$\bm e$ = [1/$e_1, \cdots, $1/$e_{MN}$]. The recursive case occurs when $K>1$, lines 4-7 partition the matrices, i.e., dividing the problem into a number of subproblems that are smaller instances of the original problem. Line 8 recursively calls {\it Diag-Blk-Inv} to compute the inverse of the sub-matrix $\bm A$. Lines 9-11 use the inverse of the sub-matrix to construct the inverse of the original matrix via the block-matrix-inversion formula. 
%

Next, we characterize the complexity of Algorithm~\ref{alg:recursive}. Let ${\cal T}(KMN)$ denote the complexity to invert a $KMN \times KMN$ matrix by the above recursive inverse procedure. In the base case, when $K=1$, we perform just the element-wise division to a vector length $MN$ in line 2, and thus ${\cal T}(MN)=\mathcal{O}(MN)$. The recursive case occurs when $K>1$. Partitioning the matrix in lines 4-7 has complexity of $\mathcal{O}(1)$. In line 8, we recursively call {\it Diag-Blk-Inv} once to invert a $(K-1)MN \times (K-1)MN$ matrix, thereby contributing ${\cal T}((K-1)MN)$  in  the overall computation. Line 9 and line 11 involve matrix multiplication, but fortunately, all the matrices  have the same special form as $\bm E$, i.e., every matrix can be divided into blocks, each of which is diagonal with dimension  $MN\times MN$. Thus, the matrix multiplications and  additions can be performed in an elementwise manner. In light of this,  the complexity of executing  line 9 and line 11 is $\mathcal{O}((K-1)^2MN)$. The  complexity of executing line 10 is the same as that in line 2, i.e., $\mathcal{O}(MN)$. Therefore, the recurrence complexity for  executing {\it Diag-Blk-Inv} is given by
\begin{equation}\label{eq: com_analysis}
\begin{aligned}
&{\cal T}(KMN)   = \\
& \left \{\begin{array}{l}
\mathcal{O}(MN) ,\hspace{4.5cm} K = 1, \\
T((K-1)MN) +\mathcal{O}((K-1)^2MN),~~ K > 1. \end{array} \right. 
\end{aligned} \tag{67}
\end{equation}
Finally, we apply the recursion-tree method~\cite{Cormen} to solve the recurrence ${\cal T}(KMN) = {\cal T}((K-1)MN) +\mathcal{O}((K-1)^2MN) ={\cal  T}((K-1)MN) + c (K-1)^2MN$, where $c$ is the implied constant coefficient. Fig.~\ref{fig:recurrence_tree} shows how we derive the recursion tree for ${\cal T}(KMN) = {\cal T}((K-1)MN) + c (K-1)^2MN$. Part (a) of the figure shows ${\cal T}(KMN)$, which progressively expands in (b)--(d) to form the recursion tree. The fully expanded tree is shown in part (d). Now we add up the complexities over all levels to get the total complexity for the entire tree as
\begin{align*}
{\cal T}(KMN) & = c (K-1)^2MN + c (K-2)^2MN + \\
& \qquad \cdots +c 2^2 MN +{\cal T}(MN) \\
& =  c (K-1)^2MN+ c (K-2)^2MN  + \\
& \qquad \cdots +c 2^2 MN + c MN\\
&= c MN \sum_{k=1}^{K-1} k^2\\
& = c MN \frac{(K-1)K(2K-1)}{6}  \\
& = \mathcal{O}(K^3 MN).
\end{align*}
Thus, the computational complexity of the recursive matrix inversion of Algorithm ~\ref{alg:recursive} is $\mathcal{O}(K^3 MN)$.

\begin{algorithm} 
	\setcounter{algorithm}{3}
	\caption{$\bm E^{-1}=$Diag-Blk-Inv$(\bm E,K,M,N)$ }  \label{alg:recursive} 
	\begin{algorithmic}[1]  
		\If {$K=1$} 
		\State $\bm E^{-1}= \text{Diag}(1./\text{Diag}(\bm E))$ 
		\EndIf
		\State $\bm A = [\bm E]_{1:(K-1)MN,1:(K-1)MN}\in \mathbb{C}^{(K-1)MN\times (K-1)MN}$ 
		\State $\bm B = [\bm E]_{1:(K-1)MN,(K-1)MN+1:KMN}\in \mathbb{C}^{(K-1)MN\times MN}$
		\State $\bm C = [\bm E]_{(K-1)MN+1:KMN,1:(K-1)MN}\in \mathbb{C}^{MN\times (K-1)MN}$
		\State $\bm D = [\bm E]_{(K-1)MN+1:KMN,(K-1)MN+1:KMN}\in \mathbb{C}^{MN\times MN}$
		\State $\bm A^{-1}=$Diag-Blk-Inv$(\bm A,K-1,M,N)$ 
		\State $\bm S = (\bm D-\bm C \bm A^{-1} \bm B)$
		\State $\bm S^{-1}=$Diag-Blk-Inv$(\bm S,1,M,N)$ 
		\State Construct the inverse matrix of $\bm E$ using the blockwise matrix inversion formula:
		\begin{align*}	
		\bm E^{-1} =\begin{bmatrix}	
		\bm A^{-1}+ \bm A^{-1} \bm B \bm S^{-1} \bm C \bm A^{-1}& -\bm A^{-1}\bm B \bm S^{-1}\\		
		-\bm S^{-1} \bm B \bm A^{-1}& \bm S^{-1}	
		\end{bmatrix}
		\end{align*}
	\end{algorithmic}
\end{algorithm}

\begin{figure}[!h]
	\setcounter{figure}{12}
	\centerline{\resizebox{.45\textwidth}{!}{\includegraphics{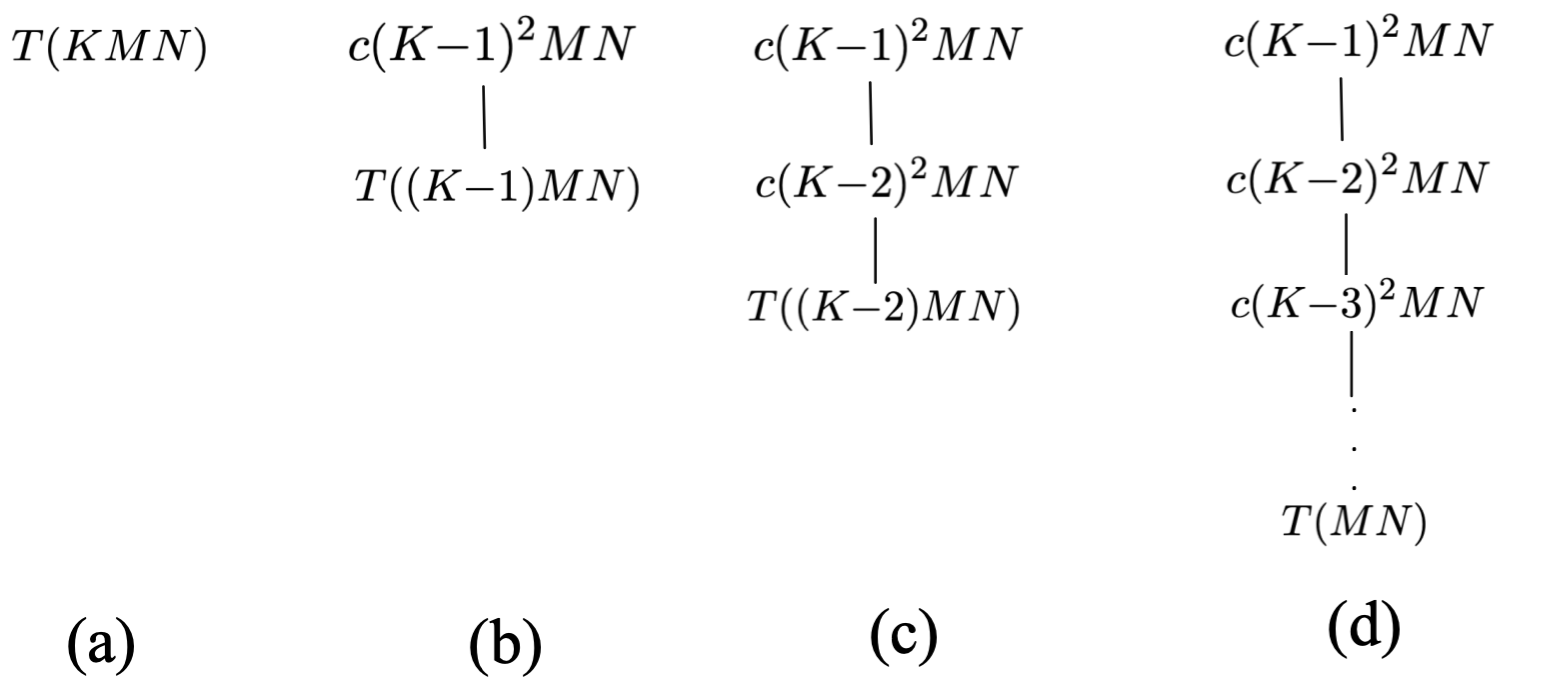}}
	}   \caption{Constructing a recursion tree for the recurrence in~\eqref{eq: com_analysis}} \label{fig:recurrence_tree}
\end{figure}

\end{document}